\documentclass[11pt]{article}

\csname @addtoreset\endcsname{equation}{section}
\let\la=\label

\def\type#1#2#3{$(\,#1\,|\,#2\,;\,\mbox{#3}\,)$}

\def\ta{ \begin{table}[h!] \begin{center} \begin{tabular}{|c|c|c|c|}
\hline &&&\\ $p$ & $(\,D\,|\,N\,)$ & $(\,d\,|\,n\,)$ & Worldvolume
multiplet \\ &&&\\ \hline &&&\\ $9$
&\type{11}{32}{M}&\type{10}{16}{MW}&Frozen\\ &&&\\ $5$ &
\type{7}{16}{SM}&\type{6}{8}{SMW}&Tensor (on-shell)\\ &&&\\ $3$ &
\type{5}{8}{PSM}&\type{4}{4}{M}& Linear/Scalar (off-shell)\\ &&&\\ $2$ &
\type{4}{4}{M}&\type{3}{2}{M}& Scalar (off-shell)\\ &&&\\ $1$ &
\type{3}{2}{M}&\type{2}{1}{MW}&Scalar (off-shell)\\ &&&\\ \hline \end{tabular}
\end{center} \caption{\small $p$-branes in $D=p+2$ dimensions.
$(D|N)$ and $(d|n)$ denote the superdimensions of the target and
worldvolume superspaces, respectively. The spinor types are denoted
by M for Majorana, W for Weyl, S for symplectic and P for pseudo. }
\la{table} \end{table}}

\def\a{\alpha}
\def\b{\beta}
\def\c{\gamma}\def\C{\Gamma}
\def\d{\delta}\def\D{\Delta}
\def\e{\epsilon}
\def\F{\Phi}
\def\h{\eta}

\def\k{\kappa}
\def\l{\lambda}\def\L{\Lambda}
\def\m{\mu}

\def\s{\sigma}

\def\th{\theta}\def\Th{\Theta}

\def\x{\xi}
\def\o{\omega}\def\O{\Omega}

\def\del{\partial}
\def\/{\over}
\def\cF{{\cal F}}
\def\cH{{\cal H}}
\def\cA{{\cal A}}
\def\cL{{\cal L}}
\def\cM{{\cal M}}
\def\cK{{\cal K}}

\def\ucM{\underline {{\cal M}}}

\def\ra{\rightarrow} 

\def\ua{\underline{\alpha}}
\def\ub{\underline{\phantom{\alpha}}\!\!\!\beta}

\def\una{\underline a}\def\unA{\underline A}
\def\unb{\underline b}\def\unB{\underline B}
\def\unc{\underline c}\def\unC{\underline C}

\def\unE{\underline E}

\def\unm{\underline m}\def\unM{\underline M}

\def\unH{\underline{H}}

\def\umu{{\underline \mu}}

\def\nab{\nabla}

\def\cE{{\cal E}}

\def\tr{{\rm tr\ }}

\def\be{\begin{equation}}
\def\ee{\end{equation}}
\def\bea{\begin{eqnarray}}
\def\eea{\end{eqnarray}}
\newcommand{\eq}[1]{(\ref{#1})}
\def\eqs#1#2{(\ref{#1}-\ref{#2})}
\def\ba{\begin{array}}
\def\ea{\end{array}}

\def\nn{\nonumber}
\newcommand{\w}[1]{\\[0.#1cm]}

\let\bm=\bibitem

\def\bd{\begin{document}}
\def\ed{\end{document}}
\def\ft#1#2{{\textstyle{{\scriptstyle #1}\over {\scriptstyle #2}}}}
\def\fft#1#2{{#1 \over #2}}

\def\se{\;\;=\;\;}
\newcommand{\mx}[4]{\left#1\begin{array}{#2}#3\end{array}\right#4}

\newcommand{\ho}[1]{$\, ^{#1}$}
\newcommand{\hoch}[1]{$\, ^{#1}$}
\newcommand{\tamphys}{\it\small Center for Theoretical Physics,
Texas A\&M University, College Station, TX 77843, USA}
\newcommand{\kings}{\it\small Department of Mathematics, King's College,
London, UK}
\newcommand{\groningen}{\it \small Institute for Theoretical
Physics, Nijenborgh 4, 9747 AG Groningen,The Netherlands}

\newcommand{\auth}{\large P.S. Howe\hoch{1}, A. Kaya\hoch{2},
E. Sezgin\hoch{2\dagger} and P. Sundell\hoch{3} }

\begin{document}

\hfill{KCL-TH-99-47}

\hfill{CTP TAMU-49/99}

\hfill{hep-th/0001169}

\hfill{\today}

\vspace{20pt}

\begin{center}
{\Large{\bf Codimension One Branes}}
\vspace{30pt}

\auth

\vspace{15pt}

\begin{itemize}

\item [$^1$] \kings
\item [$^2$] \tamphys
\item [$^3$] \groningen

\end{itemize}

\vspace{60pt}

{\bf Abstract}

\end{center}

We study codimension one branes, i.e. $p$-branes in
$(p+2)$-dimensions, in the superembedding approach for the cases
where the worldvolume superspace is embedded in a minimal target
superspace with half supersymmetry breaking. This singles out the
cases $p=1,2,3,5,9$. For $p=3,5,9$ the superembedding geometry
naturally involves a fundamental super $2$-form potential on the
worldvolume whose generalised field strength obeys a constraint
deducible from considering an open supermembrane ending on the
$p$-brane. This constraint, together with the embedding constraint,
puts the system on-shell for $p=5$ but overconstrains the $9$-brane
in $D=11$ such that the Goldstone superfield is frozen. For $p=3$
these two constraints give rise to an off-shell linear multiplet on
the worldvolume. An alternative formulation of this case is given in
which the linear multiplet is dualised to an off-shell scalar
multiplet. Actions are constructed for both cases and are shown to
give equivalent equations of motion. After gauge fixing a local
$Sp(1)$ symmetry associated with shifts in the $Sp(1)_R$ Goldstone
modes, we find that the auxiliary fields in the scalar multiplet
parametrise a two-sphere. For completeness we also discuss briefly
the cases $p=1,2$ where the equations of motion (for off-shell
multiplets) are obtained from an action principle.

{\vfill\leftline{}\vfill \vskip 10pt \footnoterule {\footnotesize
\hoch{\dagger} Research supported in part by NSF Grant PHY-9722090.
\vskip -12pt}

\pagebreak

\section{Introduction\label{intro}}

The superembedding formalism provides a powerful and systematic
method for deriving the dynamics of super $p$-branes and their
interactions \cite{se1,se2,hs1}. In this approach both the
worldvolume and target space are superspaces. Restricting the
embedding such that the fermionic worldvolume tangent space is a
subspace of the fermionic target tangent space gives rise to
equations which determine the structure of the worldvolume
supermultiplet of the $p$-brane and which may also determine the
dynamics of the brane itself.

There are several types of worldvolume supermultiplets that can
arise including scalar multiplets, vector multiplets, tensor
multiplets which have 2-form gauge fields with self-dual field
strengths, and multiplets with rank 2 or higher antisymmetric tensor
gauge fields whose field strengths are not self-dual \cite{hs1}.
Most of these cases have been studied in the superembedding
approach, primarily for embeddings with (bosonic) codimension
greater than one. In this paper we focus on codimension one
embeddings for which the basic embedding constraint gives rise to an
unconstrained scalar Goldstone superfield.

We shall consider target spaces with the minimal possible number of
supersymmetries and embeddings which preserve half the
supersymmetry. This leads to $p$-branes in $D=p+2$ dimensions for
$p=1,2,3,5,9$, as shown in Table \ref{table}. The cases of
$p=4,6,7,8$ are excluded since in these cases the minimal
worldvolume spinors have the same dimension as the minimal target
space spinors and consequently half-supersymmetry breaking is not
possible (embedding without supersymmetry breaking would imply that
the Goldstone fermions vanish and the Goldstone bosons are constant
in a physical gauge).

\ta

The basic embedding constraint mentioned above describes a relation
between the supervielbeins in the worldvolume and the target space
(see \cite{hsw1,es1,ds} for reviews). When applied to the M2 and
M5-branes in $D=11$ and to some of the D-branes in $=10$, the
embedding constraint alone yields all the dynamics. The situation is
different for the codimension one embeddings. In that case, the
embedding constraint leads to an unconstrained real scalar
superfield in a physical gauge. For the cases $p=1,2$, this
superfield is an off-shell scalar multiplet and the dynamics can be
obtained from an action principle \cite{hrs}.

For $p=3$ an additional constraint is required to obtain an
off-shell multiplet and this can in fact be done in two ways
resulting in either a linear multiplet \cite{rudy} or a dual scalar
(chiral) multiplet. The required constraint for the linear multiplet
can be deduced from considering a membrane ending on the 3-brane and
leads to a constraint on a modified 3-form field strength $\cF_3$
\cite{cs1,chsw}. For the scalar multiplet one instead has a
constraint on a modified 1-form field strength $\cF_1$ which arises
from considering a particle on the 3-brane. From either multiplet
one can then derive the dynamics from an action principle.

In this paper we shall find that the 3-brane in $D=5$ exhibits novel
features related to the existence of a triplet of 3-brane charges in
the underlying spacetime superalgebra. Both linear and scalar
formulations yield $Sp(1)_R$ covariant Green-Schwarz actions, which
are related by worldvolume dualisation of the associated $\cA_2$ and
$\cA_0$ potentials requiring a duality transformation of the
spacetime potentials. While $Sp(1)_R$-symmetry is completely broken,
the associated Goldstone scalars can be gauged away by a local
$Sp(1)$ symmetry. As a result the scalar multiplet auxiliary fields
are found to parametrise a two-sphere. In this paper we also study
the 3-brane in $D=6$ since it provides a better understanding of the
local $Sp(1)$ symmetry and, moreover, it yields the scalar multiplet
formulation of the 3-brane in $D=5$ upon vertical reduction.

For $p=5$ the additional $\cF_3$-constraint is similar to that of
the linear multiplet and indeed occurs in the M5-brane where,
however, it follows from the basic embedding condition
\cite{hs2,hsw2}}. This constraint gives the dynamics of the
worldvolume $d=6, (1,0)$ tensor multiplet directly without the use
of an action principle. The dynamics of this brane was obtained in
the superembedding formalism in \cite{ch} by imposing an additional
constraint on the embedding matrix. As we show here, this constraint
arises naturally from the geometrical $\cF_3$-constraint discussed
above. We also give the embedding of the chiral 5-brane theory in a
non-chiral theory where the equations of motion follow from an
action that involves an unconstrained 2-form potential, upon the
imposition of a non-linear self-duality condition \cite{cns}.

In the case of $p=9$, the analogous $\cF_3$-constraint turns out to
be much stronger; in fact, it freezes the worldvolume multiplet
degrees of freedom. We expect, however, that suitable modifications
of this constraint will lead to either the Horava-Witten type
9-brane \cite{hw} or the massive 9-brane \cite{bs}.

The outline of the paper is as follows: In Section \ref{emb}, we
will describe the embedding constraint and its general consequences
that follow from the use of the torsion Bianchi identities. In
Section \ref{cfc}, we will discuss the worldvolume super $3$-form,
its Bianchi identity and the action formalism. In Section
\ref{1in3} we review the cases of 1-brane in $D=3$
and 2-brane in $D=4$. In Section \ref{3in5} we study the aspects
3-brane: the six-dimensional theory, its vertical reduction to
scalar 3-brane in $D=5$ and its dual linear multiplet formulation.
In Section \ref{5in7} we show how to obtain the dynamics of the
5-brane in $D=7$ from the embedding constraint and an
$\cF_3$-constraint. The 9-brane in $D=11$ is studied in Section
\ref{9in11}. Further comments on our results, and in particular on
the 9-brane, are presented in Section \ref{concl}. In Appendix A we
give our spinor conventions and in Appendix B we collect the results
of the dimension half and up analysis of the torsion equations.


\section{The Embedding Constraint and the Torsion
Equations\label{emb}}


A $p$-brane with $n$ real supersymmetries has a $d=(p+1)$-dimensional
worldvolume moving in a $D$-dimensional spacetime with $N$ real
supersymmetries; it is described by the embedding $f:\cM\mapsto \ucM$
of a $(d|n)$-dimensional worldvolume supermanifold $\cM$ into a
$(D|N)$-dimensional spacetime supermanifold $\ucM$. We parametrise
$\cM$ and $\ucM$ with supercoordinates $z^{M}=(x^{m},\th^{\m})$, where
$m=0,...,p$ and $\m=1,...,n$, and $z^{\unM}= (x^{\unm},\th^{\umu})$,
where $\unm = 0,...,D-1$ and $\umu = 1,...,N$. The embedding matrix
$E_{A}{}^{\unA}$ is defined as

\be
E_{A}{}^{\unA} \;\;\equiv\;\;
E_{A}{}^{M}\partial_{M}z^{\unM}E_{\unM}{}^{\unA}\ ,
\la{embm}
\ee

where $E^A=dz^ME_{M}{}^{A}$ and
$E^{\unA}=dz^{\unM}E_{\unM}{}^{\unA}$ are supervielbeins in $M$ and
$\ucM$ and the 'flat' indices $A=(a,\a)$ and $\unA=(\una,\ua)$ has
the same ranges as the curved indices. By definition, the
worldvolume supervielbein is not an independent worldvolume field,
but rather induced by the embedding. (More specifically, the
embedding defines normal and tangential subspaces of the spacetime
cotangent space and tangent space, respectively. Forming an
`adapted' cotangent frame
$(\hat{E}^a,\hat{E}^{a'},\hat{E}^\a,\hat{E}^{\a'})$ for the
spacetime, such that $f^{\star}\hat{E}^{A'}=0$, $A'=(a',\a')$, we
define an induced worldvolume frame by $E^A\equiv
f^{\star}\hat{E}^{A}$.)

The basic embedding constraint is

\be
E_{\a}{}^{\una}\se 0\ .
\la{ec}
\ee

This is a natural geometric condition which states that, at each point
on the brane, the odd tangent space of the brane sits inside the odd
tangent space of the target space. For a general superembedding
preserving half-supersymmetry this constraint leads to on-shell,
off-shell or underconstrained worldvolume supermultiplets. It also
places constraints on the target space torsion (via integrability) and
is intimately related to the characteristic form of kappa-symmetry
transformations in the Green-Schwarz formalism.

In the case of codimension one embeddings listed in Table 1 one can
deduce that the embedding constraint gives in fact an unconstrained
Goldstone superfield. This can be made clear by analysing the
constraint at the linearised level. In a flat target space, and in
the physical gauge

\bea
x^{\una}&=& \big(x^a, x^{\perp}(x,\th)\big)
\nn\w2
\th^{\ua}&=&\big(\th^{\a}, \Th^{\a'}(x,\th)\big)\ .
\la{pg}
\eea

The embedding matrix \eq{embm} takes the form

\bea
&& E_\a{}^{\una}\se(0, D_\a \Phi
-i(\C^{\perp})_{\a\b'}\Th^{\b'})\ , \quad\quad E_a{}^{\unb} \se
(\d_a{}^b{},{}\del_a \Phi)\ ,
\nn\w4
&& E_{\a}{}^{\ub} \se
(\d_{\a}{}^{\b}{},{}D_\a \Th^{\b'})\ ,
\quad\quad\quad\quad\quad\quad E_a{}^{\ub} \se ({}0{},{}\del_a
\Th^{\b'})\ ,
\la{lin}
\eea

where $\Phi= x^{\perp} +\frac{i}2 \th^{\a}(\C^{\perp})_{\a\b'}
\Th^{\b'}$ and $D_{\a}$ is the flat worldvolume supercovariant
derivative. Thus the linearised form of \eq{ec} is

\be
D_\a \Phi \se i(\C^{\perp})_{\a\b'}\Th^{\b'}\ .
\la{np}
\ee

Since the matrix appearing on the right side is non-degenerate, this
equation shows that the scalar superfield $\Phi$ is unconstrained, i.e.
the content of the equation is simply to give the transverse fermionic
superfield $\Th^{\a'}$ in terms of the unconstrained transverse scalar
superfield $\Phi$.

The embedding constraint therefore yields an off-shell scalar
multiplet in the case of $p=1,2$, while it yields an
underconstrained multiplet in the case $p=3,5$. In the latter two
cases it is therefore necessary to impose further constraints, as
will be discussed in section \ref{cfc}. In ten-dimensional spacetime
an unconstrained superfield $\Phi$ (of dimension zero) can be used
to describe the $N=1$, $d=10$ off-shell supergravity multiplet. In
the case of superembedding, however, the superfield $\Phi$ (which
now has dimension $-1$) describes an induced geometry. In flat
$D=11$ superspace we will show that the induced geometry is also
flat \ref{9in11}.

In the non-linear case, the basic structure of the worldvolume
multiplets remain the same as in the linearised case, but there will
be a non-trivial geometry induced on the brane. The consequences of
the embedding constraint can be studied systematically (i.e. level
by level in the engineering dimension) by analysing the pull-back
onto the worldvolume of the target space torsion equation
$DE^{\unA}=T^{\unA}$:

\be
\nab_{A}E_{B}{}^{\unC} - (-1)^{AB}\nab_{B}E_{A}{}^{\unC} +
T_{AB}{}^{C}E_{C}{}^{\unC}\se
(-1)^{A(B+\unB)}E_{B}{}^{\unB}E_{A}{}^{\unA}T_{\unA\unB}{}^{\unC}\ ,
\la{tors}
\ee

where $T^A$ is the induced worldvolume torsion. The covariant
derivative $D$ on the target space contains the Lorentz connection
one-form $\O_{\unA}{}^{\unB}$ and the covariant derivative $\nab$ in
the worldvolume involves the pull-back of this connection and an
induced worldvolume connection one-form $\O_A{}^B$ taking values in
the Lie algebra of the unbroken worldvolume symmetries. In the case
of co-dimension embeddings $\O_a{}^b$ is $SO(p,1)$ valued.

In order to insert the embedding constraint \eq{ec} into \eq{tors} we
first parametrise the embedding matrix $E_A{}^{\unA}$ using the vector
and spinor representations of the coset element in $SO(D-1,1)/SO(p,1)$
defined by the embedding. Using the freedom in inducing the worldvolume
supervielbein $E^A$ we can set

\bea
E_a{}^{\una}&=& u_a{}^{\una}\ ,
\la{emp1}\w2
E_\a{}^{\ua}&=& u_\a{}^{\ua} + h_\a{}^{\a'}u_{\a'}{}^{\ua}\ ,
\la{emp2}\w2
E_a{}^{\ua}&=& \L_a{}^{\a'}u_{\a'}{}^{\ua}\ ,\la{emp3}
\eea

where the primed indices label the normal directions in the super
tangent bundle of the target superspace. The matrices $u_{\a}{}^{\ua}$
and $u_{\a'}{}^{\ua}$ forms the coset representative in
$Spin(D-1,1)/Spin(p,1)$, and the matrix $u_{a}{}^{\una}$ together with
a vector $u_{\perp}{}^{\una}$ make up the corresponding representative
in $SO(D-1,1)/SO(p,1)$. In the cases of $p=3,5$, where the target space
group includes an internal symmetry factor, it will be understood that
the meaning of the spinorial $u$-matrix is modified appropriately. The
normal embedding matrix can be chosen to be:

\be
\ba{lclclcl}
E_{\perp}{}^{\unb}& = &u_{\perp}{}^{\unb} &,&E_{\perp}{}^{\ub} &= &0
\w2
E_{\a'}{}^{\una} &=& 0 &,& E_{\a'}{}^{\ub}&=&u_{\a'}{}^{\ub} \\
\ea
\ee

The inverses $(E^{-1})_{\unA}{}^A$ and $(E^{-1})_{\unA}{}^{A'}$ are
given by

\be
\ba{lclclcl}
(E^{-1})_{\una}{}^{b} &=&u_{\una}{}^{b} &,&
(E^{-1})_{\una}{}^{\b} &=& 0
\w2
(E^{-1})_{\ua}{}^{b} &=&0 &,&
(E^{-1})_{\ua}{}^{\b} &=& u_{\ua}{}^{\b} \ea \ .\la{vb}
\ee

and

\be
\ba{lclclcl}
(E^{-1})_{\una}{}^{\perp}&=& u_{\una}{}^{\perp} &,&
(E^{-1})_{\una}{}^{\b'}&=& -u_{\una}{}^{a}\L_{a}{}^{\a'}
\w2
(E^{-1})_{\ua}{}^{\perp} &=& 0 &,& (E^{-1})_{\ua}{}^{\b'} &=&
u_{\ua}{}^{\b'}-u_{\ua}{}^{\a}h_{\a}{}^{\b'}
\ea
\ee

Notice that the superfields $u$, $h_\a{}^{\a'}$ and $\L_a{}^{\a'}$
can be expressed explicitly in terms of the fundamental embedding
superfields $z^{\unM}(z^M)$ leading to a formulation of the
embedding as non-linear supersymmetric sigma-model. The
parametrisation in \eq{emp3} turns out to be more convenient,
however, in the study of the non-linearities of the superembedding.

There are two equivalent ways to induce the worldvolume $SO(p,1)$
connection $\O_A{}^B$. The first one is to fix some of the
components of the worldvolume torsion $T_{AB}{}^C$ in a convenient
form. The second one is to choose $\O_A{}^B$ such that the
$SO(D-1,1)$ valued worldvolume covariant derivatives of the Lorentz
harmonics defined by

\be
X_A\;\;\equiv\;\;(\nabla_A u)u^{-1}
\la{defx}
\ee

(the contraction is over the underlined index) satisfy

\bea
X_{A,B}{}^{C} &=& (\nabla_A u_B{}^{\unC}) (u^{-1})_{\unC}{}^C \se 0
\ ,\nn\w3
X_{A,B'}{}^{C'} &=& (\nabla_A u_{B'}{}^{\unC})
(u^{-1})_{\unC}{}^{C'} \se 0 \ .
\la{indx}
\eea

Using this method, which turns out to be more efficient, the
worldvolume torsion $T_{AB}{}^C$ can be computed in terms of the local
composite connection $X_{A,B'}{}^C$, the embedding matrix components
$h_\a{}^{\b'}$ and $\L_a{}^{\b'}$ and the pull-back of the target space
torsion. Notice that once the torsion equation \eq{tors} and the target
space torsion Bianchi identity $DT^{\unA}=E^{\unB}R_{\unB}{}^{\unA}$
have been solved, the worldvolume torsion Bianchi identity $\nab
T^A=E^B R_B{}^A$ is identically satisfied.

To proceed further, we consider target space superspaces with dimension
$0$ torsion components

\be
T_{\ua\ub}{}^{\una}\se -i(\C^{\una})_{\ua\ub}\ .
\la{tt}
\ee

By systematically analysing the effect of the embedding constraint
\eq{ec} in the torsion equation \eq{tors}, one can then determine all
of the components of the induced worldvolume geometry as well as the
covariant form of the non-linear worldvolume scalar multiplet. For our
main purposes, however, we shall only need the dimension zero results.
The results of the higher dimensional components are collected in
Appendix B.

The dimension zero component of \eq{tors}, given the embedding
condition \eq{ec}, is

\be
E_{\a}{}^{\ua} E_{\b}{}^{\ub} T_{\ua\ub}{}^{\unc} \se
T_{\a\b}{}^{c} E_{c}{}^{\unc}\ .
\la{tors0}
\ee

Right-multiplying this equation with the tangential components
$(E^{-1})_{\unc}{}^c$ of the inverse embedding matrix we find

\be
T_{\a\b}{}^c \se -i\left(
(\C^c)_{\a\b}+h_\a{}^{\a'}h_\b{}^{\b'}(\C^c)_{\a'\b'}\right)\ ,
\ee

while the normal component yields an algebraic condition on the following component
of the spinor-spinor part of the embedding matrix:

\be
h_{(\a}{}^{\a'}(\C^{\perp})_{\b)\a'}\se 0\ .
\ee

It is convenient to define

\be
h_{\a\b}\;\equiv\; h_{\a}{}^{\b'}(\C^{\perp})_{\b'\b}\ ,
\ee

such that the above two equations can be written as

\bea
T_{\a\b}{}^{c} &=& -i\left(
(\C^{c})_{\a\b}-h_{\a\c}(\C^c)^{\c\d}h_{\d\b}\right)
\la{T}\w2
h_{(\a\b)}&=& 0\ . \la{h}
\eea

\section{The ${\cF}$-Constraints and the Action Formula\label{cfc}}

\subsection{Closed Forms and Cartan Integrable Systems\label{cis}}

The required new ingredient for $p=3,5,9$ is a constrained, {\it
globally} defined generalised super three-form field strength
$\cF_3$ (or, alternatively, $\cF_1$ in the case $p=3$). Forms of
this type can be introduced in a natural geometrical fashion by
considering open superbranes ending on other superbranes as
discussed in \cite{cs1,chsw}. As will be explained in Section
\ref{action}, the existence of the $\cF_3$-forms follows from the
basic requirements of configurations where open membranes (or, for
$p=3$, particles) end on the codimension one branes.

We shall begin, however, by using an alternative method based on the
construction of Wess-Zumino forms as the pull-backs of target space
Cartan Integrable Systems, i.e. a collection of forms obeying
generalised Bianchi identities. For the cases we are considering these
systems are:

\be
\ba{lcrclrclrcl}
D=7,11 &:& dH_7 &=& {1\over 2}H_4H_4\ ,& dH_4 &=& 0\ , &&
\w3
D=5   &:& dH_5^r &=& H_4^rH_2\ ,& dH_4^r &=& 0\ , & dH_2=0\ ,\ \ r=1,2,3\ ,&
\w3
D=4   &:& dH_4 &=& 0\ , &&&&&
\w3
D=3   &:& dH_3 &=&0\ ,  &&&&&
\ea
\la{dh}
\ee

where we have suppressed the wedge product notation. These Bianchi
identities arise in the superspace formulation of the relevant
target space supergravities (possibly coupled to matter). They take
the same form in the limit of flat target superspace as well,
although the details of the solutions to them, of course, simplifies
in that limit. The triplet of closed five-forms and four-forms for
$D=5$ can be viewed as arising from the triplet of closed five-forms
in $N=(1,0), D=6$ superspace. These correspond to the triplet of
tensorial charges which are allowed in the $N=(1,0), D=6$
super-Poincar\'{e} algebra.

The  equations \eq{dh} imply that locally we can write

\be
\ba{lcrclrclrcl}
D=7,11 &:& H_7 &=& dC_6+{1\over 2}C_3H_4\ , & H_4 &=& dC_3\ , &&
\w3
D=5   &:& H_5^r &=& dC_4^r(t)+(1-t)C_1H_4^r+tC_3^rH_2\ ,& H_4^r &=& dC_3^r\ ,
 & H_2=dC_1\ ,&
\w3
D=4   &:& H_4 &=& dC_3\ , &&&&&
\w3
D=3   &:& H_3 &=&dC_2 \ ,  &&&&&
\ea
\la{hb}
\ee

where $t$ is an arbitrary constant and

\be
C_4^r(t)\se C^r_4+tC_1C^r_3\ .
\la{ct}
\ee

The idea now is to construct the globally well-defined Wess-Zumino
form $W_{p+2}$ in the worldvolume, in terms of the pull-backs of
these target space superforms. In addition to being globally
defined, this form must be closed

\be
dW_{p+2} \se 0\ ,
\ee

and this implies that locally we can write

\be
W_{p+2} \se dZ_{p+1}\ .
\la{z}
\ee

As we shall see in the next section, the form $Z_{p+1}$ gives the
Wess-Zumino term in the Green-Schwarz formalism when restricted to the
bosonic part of the worldvolume.

For $p=1,2$, the Wess-Zumino forms can be readily constructed as the
pull-backs of $H_3$ and $H_4$, respectively:

\be
\ba{lcrclrclrcl}
p=1 &:& W_3 &=& {\unH}_3\ , &&&&&
\w3
p=2  &:& W_4 &=& {\unH}_4\ , &&&&&
\ea
\la{p12}
\ee

where the underlining of the target superforms denote their pull-backs
to the worldvolume. However, for $p=3,5,9$ we need to introduce a set
of globally well-defined forms satisfying the following Bianchi
identities

\bea d{\cF}_3 &=& {\unH}_4\ , \quad\quad p=3,5,9\ , \la{dfh}\w2
d{\cF}_1 &=& {\unH}_2\ , \quad\quad p=3\ , \la{dfh2} \eea

which implies that locally

\bea
{\cF}_3 &=& d\cA_2 + {\unC}_3 ,
\la{fbi3}\w2
{\cF}_1 &=& d\cA_0 + {\unC}_1\ ,
\la{fbi3b}
\eea

where $\cA_0$ and $\cA_2$ are worldvolume superforms. Note that for
$p=3$ we take a linear combination of the triplet of closed
four-forms $H_4^r$, i.e.

\be
\unH_4\se Q^r \unH_4^r\ ;\qquad \unC_3\se Q^r \unC_3^r
\la{fbi3c}
\ee

where $\vec Q$ is a three-vector of real constants. These constants are
related to the tensorial charge of the $N=1$, $D=5$ superalgebra
carried by the $3$-brane. Equipped with ${\cF}_3$ and ${\cF}_1$, we can
construct the appropriate Wess-Zumino forms as follows:

\be
\ba{lcrclrclrcl}
p=3 &:& W_5 &=& {\unH}_5 -(1-t) {\unH}_2\,{\cF}_3 - t {\unH}_4\, {\cF}_1\ , &&&&&
\w3
p=5  &:& W_7 &=& {\unH}_7- {1\over 2}{\unH}_4\,{\cF}_3\ , &&&&&
\w3
p=9 &:&  W_{11}&=& {\unH}_4 \big(-{\unH}_7+ \ft12 {\unH}_4\, {\cF}_3\big) \ .&&&&&
\ea
\la{p359}
\ee

The forms $Z_{p+1}$ defined by $W_{p+2}=dZ_{p+1}$ can be chosen to be

\be
\ba{lcrclrclrcl}
p=1&:& Z_2 &=& {\unC}_2\ , &&&&&
\w3
p=2  &:& Z_3 &=& {\unC}_3\ , &&&&&
\w3
p=3 &:&  Z_4&=& {\unC}_4(t) +(1-t) {\unC}_1{\cF}_3+t {\unC}_3 {\cF}_1\, &&&&&
\w3
p=5 &:&  Z_6&=& {\unC}_6 +{1\over 2}{\unC}_3 {\cF}_3\ , &&&&&
\w3
p=9 &:& Z_{10} &=& {\unC}_3\left( {\unH}_7- \ft12 {\cF}_3 {\unH}_4\right)\ .
\ea
\la{z12359}
\ee

Note that for $p=3$, ${\unC}_4$ and $\unH_5$ are again given as linear
combinations of $\unC_4^r$ and $\unH_5^r$,

\be
\unH_5 \se Q^r \unH_5^r;\qquad \unC_4=Q^r \unC_4^r\ .
\ee

where $Q^r$ are the same constants as the ones used in \eq{fbi3c}.

The generalised field strengths obey the following $\cF$-constraints
\cite{cs1,chsw}:

\bea
{\cF}_{\a AB} &=& 0\ ,
\la{fc}\w2
{\cF}_{\a} &=& 0\ .
\la{fc2}
\eea

The derivation of these constraints in the context of open membranes
ending on the codimension one branes is briefly explained at the end
of next section. Equations \eq{fc} and \eq{fc2} are consistent with
the following constraints on the dimension zero components of the
target space superforms

\be
D=3,4,7,11\quad :\quad H_{\underline{a_1\cdots
a_k\a\b}}\se -i(\C_{\underline{a_1\cdots a_k}})_{\underline{\a\b}}\ ,
\la{h0}
\ee

where the relevant values of $k$ can be seen from \eq{hb}, and the
spinor types are given in Table 1. Note that for $D=7$, $\ua\ra\ua i$
and $(\C_{\una\cdots})_{\underline{\a\b}} \ra (\C_{\una\cdots})_{\underline{\a\b}}
\e_{ij}$. In $D=5$ the spinors are symplectic Majorana and the dimension
zero components of $H^r_5,H^r_4,H_2$ are taken to be

\be
D=5\ :
\mx{\{}{lcl}{H^r_{\una\unb\unc\ua\ub} &=& (\C_{\una\unb\unc}\,\s^r)_{\ua\ub}
\w2
H^r_{\una\unb\ua\ub} &=&-i(\C_{\una\unb}\,\s^r)_{\ua\ub}
\w2
H_{\ua\ub} &=&-C_{5\ua\ub}}{.}
\la{hhh}
\ee

where $(\s^r)_{ij}$ are the symmetric Pauli matrices. The
constraints can be rewritten in a manifestly six-dimensionally
covariant form: $H_2$ is identified with the six-dimensional torsion
component $-T^6$ (satisfying \eq{tt}) and
$H^r_5$ and $H^r_4$ are combined in a triplet of six-dimensional
$5$-forms satisfying

\be
D=6 \quad : \qquad dH^r\se 0\ ,\qquad
H^r_{\underline{abc\a\b}}\se (\C_{\una\unb\unc}\,\s^r)_{\ua\ub}\ .
\la{h6}
\ee

The above expressions for the dimension zero components of the
$H$-forms are valid in flat superspace where all other components
vanish. In curved superspace, on the other hand, these expressions
may require some modifications depending on the supergravity theory
under consideration.

\subsection{The Action Formula\label{action}}

Since the Wess-Zumino form $W_{p+2}$ is a closed $(p+2)$-form on a
manifold which has bosonic dimension $(p+1)$ it follows that it is
exact. This is so because the de Rham cohomology of a supermanifold
coincides with the de Rham cohomology of its body. Therefore we can
always (i.e. for {\it any} embedding) write

\be
W_{p+2} \se d K_{p+1}
\la{w}
\ee

for some {\it globally} defined $(p+1)$-form $K$ on $M$. Furthermore,
since none of the target space fields or the worldsurface fields has
negative dimension, at least for the models under discussion here, it
follows that the only non-vanishing component of $K$ is the purely
bosonic one, i.e.

\be
K_{\a A_1 \cdots A_p} \se 0\ .
\la{kc}
\ee

We remind that once $K_{a_1\cdots a_{p+1}}$ has been obtained from the
dimension zero component of \eq{w} (with indices $\a\b a_1\cdots a_p$)
the remaining dimension half and one components of \eq{w} are
identically satisfied.

We now define the Green-Schwarz Lagrangian form $L_{p+1}$ to be
\cite{hrs}

\be
L_{p+1} \se K_{p+1}-Z_{p+1} ,
\la{L}
\ee

with $Z_{p+1}$ defined in \eq{z}. In view of \eq{z} and \eq{w}, we have

\be
dL_{p+1}=0\ .
\ee

Under a worldsurface superdiffeomorphism generated by the vector field
$v$ one has

\be
\d L_{p+1} \se d i_v L_{p+1}\ .
\la{9a}
\ee

Therefore, the action integral

\be
S \se \int_{M_0}\, L^0_{p+1}\ ,
\la{9b}
\ee

where $M_0$ is the body of $M$ and where

\be
L^0_{p+1} \se dx^{m_{p+1}}\wedge dx^{m_p} \wedge
\ldots dx^{m_1} L_{m_1\ldots m_{p+1}}\ ,
\la{9c}
\ee

will be invariant under $\k$-symmetry transformations and
diffeomorphisms of $M_0$, since these transformations are identified
with the leading components of $v$. The vertical bars, that indicate
evaluation of a (worldvolume) superfield at $\th=0$, will be dropped
in the rest of the paper.

In the case of $p=1,2,3$ the worldvolume multiplets are off-shell
and which means that the action \eq{9b} can actually be generalised
to a full superspace actions \cite{hs1}.

The super-de-Rham cohomology theorem mentioned above also explains
the role of open membranes in deriving the $\cF_3$-constraint. We
refer the reader to \cite{cs1,chsw} for a rigorous proof. The basic
idea, however, is that the cohomology theorem implies that there
exists a globally defined three-form $K_3$ on the membrane obeying
$dK_3=\unH_4$. We then define the three-form $\cF_3$ on the
codimension one brane by

\be f_1^{p\,\star}\cF_3\se f_1^{2\,\star} K_3\ , \la{f1p} \ee

where $f_1^p$ and $f_1^2$ are the embeddings of the membrane
boundary in the codimension one brane and the membrane,
respectively. By construction $\cF_3$ is globally defined and obeys
\eq{dfh}. If we now assume that the membrane obeys the embedding
constraint \eq{ec} then it can be shown that \cite{cs1,chsw} all
involved embeddings obey \eq{ec}. Moreover, it also follows (on
dimensional grounds) that $K_3$ satisfies $i_\x K_3=0$ for any
fermionic vector $\x$. The $\cF$ constraint \eq{fc} now follows by
taking the inner derivative of \eq{f1p} and noting that the
embedding condition on $f_1^p$ implies that the push-forward
$f_{1\,\star}^p\x$ is a fermionic vector that can be varied
independently in the fermionic tangent space of the codimension one
brane.

\section{String in $D=3$ and membrane in $D=4$\label{1in3}}

Let us begin our analysis of codimension one branes by reviewing the
string in three dimensions and the membrane in four dimensions.
Since results for $p=1,2$ are already available in the literature
\cite{hrs} our presentation here will be brief.

Let us consider first the string in $D=3$. A 2 component $D=3$
Majorana spinor $\psi$ splits into two Majorana-Weyl spinors
$\psi_+$ and $\psi_-$ on the superstring worldsheet. Let the
positive chirality label the tangential direction and the negative
chirality label the normal direction. Moreover, let us take the
$D=3$ target space to be flat for simplicity.

The tangential equations \eqs{T1}{T4} given in Appendix B determine
the induced torsion component in terms of various quantities. The
normal equations \eqs{A}{F}, also given in Appendix B, yield the
following results

\bea
&& h_{++} \se 0,
\nn\w2
&& X_{+,a}\se i\L_{a+}\ ,\quad\quad \nab_+\L_{a+}\se
\d_{+\!\!\!\!_{+}}^b X_{a,b}\ ,
\nn\w2
&& X_{[a,b]}\se 0 \ , \quad\quad \nab_{[a}\L_{b]}\se 0 \ .
\la{13}
\eea

These equations describe an $N=(1,0)$ off-shell scalar multiplet on
the string worldsheet. To see this more explicitly, we note that the
full content of \eq{13} at the linearised level, which can be
deduced by using \eq{pg} and \eq{lin}, is given by

\bea
D_+ \Phi &=& i\psi \ , \quad\quad D_+ \psi \se
\d_{+\!\!\!\!_{+}}^a \del_a \Phi\ ,
\nn\w2
X_{a,b}&=& \del_a \del_b \Phi \ ,\quad\quad  \L_a \se \del_a \psi \ ,
\eea

where $ \psi  := \Th^{'-} $.

The Green-Schwarz action can be obtained by using \eq{L}, \eq{9b}
and \eq{9c}. In flat target space, the only non-vanishing component
of $H_3$ is given in \eq{h0}. Next, recall the definitions

\be
W_3 \se {\unH}_3 \se \cases{
d K_2  &  \quad\quad {\rm globally} \cr
       &                            \cr
d{\unC}_2& \quad\quad {\rm locally}   }
\ee

From \eq{w} and \eq{kc} one finds that the only non-vanishing
component of $K_2$ is

\be
K_{ab}\se -\e_{ab}\ .
\ee

Using the above results in \eq{L} yields the action

\be
S_2\se\int_{M_o}\frac12 E^a E^b(\e_{ba}+ {\unC}_{ba})\se
\frac12 \int_{M_o}\frac12 d^2x(\sqrt{-\det{g}}+\e^{mn}{\unC}_{mn})\ ,
\ee

where $M_o$ is the body of the worldvolume superspace and where the
induced metric $g_{mn}$ is given in terms of the supersymmetric line element
$\cE_m{}^{\una}=e_m{}^a E_a{}^{\una}|$ by

\be
g_{mn}\se e_m{}^a
e_n{}^b\eta_{ab}\se \cE_m{}^{\una}\cE_n{}^{\unb}\h_{\una\unb} \ .
\la{gsm}
\ee

We now turn to the membrane in $D=4$. Again, we consider flat target
superspace for simplicity. In this case all the $Y$- and $Z$-tensors
vanish and from \eq{T1}, \eq{A} and \eq{B} and we readily find

\bea
h_{\a\b}&=& h C_{\a\b} \ ,
\nn\w2
\nab_{\a}h &=&-i \frac{1+h^{2}}{2} (\c^{a})_{\a}{}^{\d}\L_{a\d}\ ,
\la{sol4}  \w2
T_{\a\b}{}^{a} &=& -i(1+h^{2})(\c^a)_{\a\b} \ .
\nn
\eea

This system describes an $N=1$ off-shell scalar multiplet in $d=3$.
The field $h$ plays the role of an auxiliary field. An action is
needed to set $h=0$ which leads to the Dirac equation
$(\c^a\L_a)_\a=0$ and the remaining equations of motion \cite{hrs}.
Indeed, from \eq{p12} and \eq{h0} one finds that the only
non-vanishing component of the superform $K_3$ defined by $W_4=dK_3$
is \cite{hrs}

\be
K_{abc} \se \e_{abc}\,K\ ,
\quad\quad
K \se {1-h^2\over 1+h^2}\ .
\la{37}
\ee

and

\be
\nab_\a K \se {2ih\over 1+h^2}\, (\c^a\L_a)_\a\ .
\la{dk}
\ee

As was explained in Section \ref{action}, this result is consistent
with \eq{37} without any condition on $\L$. A Lagrangian can be
constructed by using the formula \eq{L}, which yields the result
\cite{hrs}

\be
\cL \se \sqrt{-\det g}\ \left({1-h^2\over 1+h^2} \right)
-{1\over6}\e^{mnp}{\unC}_{mnp}\ ,
\la{39}
\ee

where $g$ is again the standard GS induced metric given in \eq{gsm}.
The only difference from the usual GS Lagrangian is the presence of
the auxiliary field $h$. However, the equation of motion for this
field is purely algebraic and can be used to set $h=0$. We thus
recover the standard GS action.

\section{3-brane in $D=5$\label{3in5}}

As we have remarked previously there are two formulations of the
3-brane in $D=5$: the first has a worldvolume linear multiplet and
correspondingly a 3-form field strength $\cF_3$, while the second
has a worldvolume scalar multiplet with a 1-form field strength
$\cF_1$. In the five-dimensional context the first is perhaps more
natural since it has only one physical scalar corresponding to the
transverse direction. The scalar multiplet, on the other hand, which
has two physical scalars is more naturally formulated in a
six-dimensional context. The $3$-brane in $D=6$ was discovered some
time ago \cite{bst}, and here we shall begin by presenting its
covariant formulation with auxiliary fields, and then demonstrate
that how its vertical reduction gives the scalar formulation of the
$3$-brane in $D=5$. We shall then obtain the linear multiplet
formulation in the GS formalism by dualising the leading component
$\cF_a|$ of $\cF_1$. Finally we shall give the linear multiplet
formulation directly as a superembedding.

\subsection{A Useful Digression: The 3-brane in $D=6$\label{3in6}}

The odd-odd part of the embedding matrix $E_{\a}{}^{\una}$ requires
slightly more careful treatment for the three-brane because of the
presence of an internal symmetry group in the target space, and
because we want to split the target spinor index (which is
symplectic Majorana-Weyl) into two four-dimensional Majorana spinor
indices. As discussed in section \ref{emb}, the spinor-spinor part of the
embedding matrix can be written

\be
E_{\a}{}^{\ua} \se {1\over\sqrt{2}}\left[(u\otimes v_1)_{\a}{}^{\ua}
+(\c^5u\otimes v_2)_{\a}{}^{\ua}\right] +{1\over\sqrt{2}}h_{\a}{}^{\b}
\left[(\c^5 u\otimes v_1)_{\b}{}^{\ua} -(u\otimes v_2)_{\b}{}^{\ua}
\right]
\la{euv}
\ee

The notation here is as follows: $u$ is a $4\times 4$ matrix
belonging to the six-dimensional spin group $Spin(1,5)$, $(v_1,v_2)$
are both two-component objects which together make up an element of
$Sp(1)$ (so when $\ua\ra \ua i$ one replaces $v_1\ra v_1{}^i$,
$v_2\ra v_2{}^i$, where $v_I{}^iv_J{}^j\e_{ij}=\e_{IJ}$) and the
underlined spinor index $\ua$, running from 1 to 8, is a combined
six-dimensional Majorana-Weyl spinor $\otimes$ internal $Sp(1)$
doublet index.

The linearised analysis of the embedding constraint $E_\a{}^{\una}=0$
along the lines explained in Section \ref{emb}, in this case gives the
constraint

\be
D_\a\Phi^{a'}\se i\left(\C^{a'}\right)_{\a\b'}\Theta^{\b'}\ ,\qquad
a'=1,2\ ,
\ee

where $\Phi^{a'}=x^{a'}+\frac{i}2 \th^{\a}(\C^{a'})_{\a\b'}
\Th^{\b'}$ is the Goldstone superfield which describes an off-shell
$N=1$, $d=4$ scalar supermultiplet.

In the non-linear case, the dimension zero worldvolume torsion is
given by the standard form \eq{T}. In this case, $h_{\a\b}$
satisfies \eq{h}, which follows from the projection of \eq{tors0} on the fifth
direction, and

\be
\left(h\c^5\right)_{(\a\b)}\se 0\ ,
\la{h6b}
\ee

which follows from the projection on the sixth direction (notice
that the matrix $v_I{}^i$ drops from these calculations since the
target space torsion contains the $Sp(1)$ invariant $\e_{ij}$). The
solution to \eq{h} and \eq{h6b} is

\be
h_{\a\b}\se A C_{\a\b} + i\, B (\c^5)_{\a\b}
\se \ft12(1+\c^5)z+\ft12(1-\c^5)\bar{z}\ ,\qquad z\se A+iB\ ,
\la{hab}
\ee

where $A$ and $B$ are real worldvolume superfields. Thus, in
addition to the Goldstone superfield associated with the breaking of
the super-translation group, we have so far introduced the three
scalar superfields $v_I{}^j$ parameterizing $Sp(1)$ and the two
additional superfields $A$ and $B$.

Naively, one would expect that the leading components of $A$ and $B$
are related to the auxiliary fields of the Goldstone superfield, and
that the leading components of $v_I{}^i$ are the Goldstone
superfields associated with the complete breaking of the
$R$-symmetry group $Sp(1)$. However, as we shall see below, there
exists an extra local $Sp(1)$ symmetry in the spinor-spinor part of
the embedding matrix, which enables us to either gauge away
$v_I{}^i$, or equivalently, to gauge away the fields $A,B$ and a
scalar in $v_I{}^i$ associated with a $U(1)$ subgroup of $Sp(1)$. In
both cases, the remaining two superfields will indeed be related to
the auxiliary fields of the Goldstone superfield and as we shall see
they parametrise the two-sphere $Sp(1)/U(1)$.

To exhibit the extra symmetry we first use a combined super-Weyl and
transverse $SO(2)$ transformation $\l$ (generated by $1$ and $i\c^5$)
of the induced worldvolume supersechsbein to replace the spinor-spinor
component of the embedding matrix by the following equivalent
parametrisation

\be
E_{\a}{}^{\ua} \se \ft1{\sqrt{2}} h^+{}_\a{}^\b\left[(u\otimes
v_1)_{\b}{}^{\ua} +(\c^5u\otimes v_2)_{\b}{}^{\ua}\right]
+\ft1{\sqrt{2}}h^-{}_{\a}{}^{\b} \left[(\c^5 u\otimes
v_1)_{\b}{}^{\ua} -(u\otimes v_2)_{\b}{}^{\ua} \right]\ ,
\la{me}
\ee

where  $h^{\pm}\se (h^1\pm h^2)/\sqrt{2}$ and

\be
h^i\se A^i+i\c^5
B^i\se \ft12(1+\c^5)z^i+\ft12(1-\c^5)\bar{z}^i\ ,\qquad
z^i\se A^i+iB^i\ .
\ee

Defining the worldvolume chiral projections $E_\pm=\frac12(1\pm
\c^5)E$, equation \eq{me} can be written in the following manifestly
$Sp(1)$ invariant form

\be
E_{+} \se z^i v_i\otimes u_+\ ,\qquad E_- \se \bar{z}^i v_i \otimes u_-\ .
\la{sp1inv}
\ee

The freedom in inducing the worldvolume supersechsbein, i.e. $E\sim \l
E$, implies the following equivalence relation between embeddings

\be
(z^1,z^2)\sim \l\, (z^1,z^2)\ ,\qquad \l\in {\bf C}\ ,\quad \l\neq 0\ .
\ee

Notice that $\l$ is a local superfield. Hence inequivalent
embedding matrices are parametrised by maps to $CP^1$. The
parametrisation in \eq{euv} and \eq{hab} corresponds to the
representative

\be
(z,1)\se (z^-/z^+,1)\ ,
\ee

where

\be
\mx{(}{c}{z^-\\z^+}{)}\se \O\mx{(}{c}{z^1\\z^2}{)}\ ,
\qquad \O \se (1-i\s^2)/\sqrt{2}\ .
\ee

In this parametrisation, the local $Sp(1)$ transformation
$v\rightarrow g\,v$ acts on $z$ by
the M\"{o}bius transformation $z\rightarrow (az+b)/(cz+d)$ such that

\be
(cz+d\,)E_+\left( g\, v\, ,{az+b\over cz+d}\right) \se E_+(v\,,z)\ ,
\ee

where $a=\bar{d}$ and $b=-\bar{c}$ make up the $Sp(1)$ matrix

\be
\mx{(}{ll}{a&b\\c&d}{)} \se \O(g^{-1})^T\O^{-1}\ .
\ee

Thus we can fix $v=1$, in which case $CP^1$ is parametrised by $z$,
or $z=0$, in which case $CP^1= Sp(1)/U(1)$ (where $U(1)$ is the
stability subgroup that leaves $z=0$ invariant) is parametrised by the
coset representative

\be
v \se \O(L^{-1})^T\O^{-1}\ ,
\ee

where

\be
L \se {1\over \sqrt{1+|\phi|^2}}\mx{(}{ll}{1&\phi\\ -\phi^*&1}{)}\ .
\ee

The right action of $U(1)$ on $v$ is generated by $\s^1$ and its
right action on $L$ is generated by $\O\s^1\O^{-1}=-\s^3$. The two
gauge choices are related by $\phi=-z$.


Let us first consider the $v=1$ gauge. In this case, the dimension
zero worldvolume torsion is given by

\be
T_{\a\b}{}^c\se -i(1+\bar{z}z)(\c^c)_{\a\b}\ .
\ee

As mentioned in Section \ref{cis}, there exists an $Sp(1)$ triplet
of closed five-forms $H^r_5$ in the $N=(1,0)$, $D=6$ superspace. Its
dimension zero components are given by \eq{h6}. To construct a
$\k$-symmetric action for the off-shell scalar (chiral)
supermultiplet along the lines described in Section \ref{action}, we
take the Wess-Zumino form in \eq{w} to be $W_5=Q^r\unH^r_5$, where
$Q^r$ is the tensorial charge of the $N=(1,0)$, $D=6$
superalgebra carried by the $3$-brane. The resulting kinetic term is
given by

\be
K_{abcd}\se \e_{abcd} K \ ,\qquad
K\se {Q_1 (1-\bar{z}z)+q\bar{z}+\bar{q}z \over 1+\bar{z}z}\ ,
\la{k6}
\ee

where we have defined

\be
q\;\;\equiv\;\; q_1+iq_2\se Q_3+i Q_2\ .
\la{q}
\ee


Let us now consider the $z=0$ gauge instead. The dimension zero
worldvolume torsion in this case is given by

\be
T_{\a\b}{}^c\se -i(\c^c)_{\a\b}\ ,
\ee

Note that the $v$'s drop out of this expression by virtue of the
fact that $\e_{ij}$ is $Sp(1)$ invariant. The kinetic term is now
found to be

\be
K\se \ft12Q^r\tr (\s^1 v\,\s^r\,v^{-1})\se \ft12
Q^r\tr (L^{-1}\s^3 L \O^{-1}\s^r\O)\ .
\ee

Introducing the $Sp(1)$ generators $[T_i,T_j]=\e_{ijk}T_k$, we define
so called $C$ and $S$ functions by

\be
L^{-1}T_3L\se C T_3 + \bar{S} T_+ + S T_-\ ,\la{cs}
\ee

where $T_\pm=\ft12(T_1\pm i T_2)$. In terms of these functions

\be
K \se Q_1C- q\bar{S}-\bar{q}S\ .
\la{kq}
\ee

The $C$ and $S$ functions are easily found from \eq{cs} to be

\be
C\se {1-|\phi|^2\over 1+|\phi|^2} \ ,\qquad
S\se {2\phi  \over 1+\bar{\phi}\phi}\ ,
\ee

which together with $\phi=-z$ shows the equivalence between \eq{kq} and the
kinetic term \eq{k6} derived in the $v=1$ gauge.

The complete action is therefore given by

\be
\cL\se  {Q_1(1-\bar{z}z)+q\bar{z}+\bar{q}z \over 1+\bar{z}z}
\sqrt{-\det{g}} - \ft1{4!}\e^{mnpq}\vec{Q}\cdot\vec{\unC}_{mnpq}\ ,
\la{l6}
\ee

where $g_{mn}$ is the induced metric in \eq{gsm}.
The Wess-Zumino term breaks the local $Sp(1)$ to a local $U(1)$ with
generator $Q^r (\s^r)_{ij}$. When this $U(1)$ coincides with the
local right action on the coset, i.e. when $[Q^r \s^r,\s^1]=0$, the
resulting single $U(1)$ is a local symmetry of the action. This
condition implies $q=0$, as can also be seen directly from \eq{l6}
by inspection.

In order to give a global description of the $CP^1$ manifold we
cover the upper hemisphere with coordinate $z$ and the lower
hemisphere with $1/z$. Therefore, we must check that the action is
invariant on the overlap region, modulo possible transformations of
other fields in the theory (including the target space fields). We
find that the action \eq{l6} is invariant under the following
combined transformations:

\be
z \ra {1\over z}\ , \qquad (Q_1,Q_2,Q_3) \ra (-Q_1,Q_2,-Q_3)
\ ,\qquad (C_4^1,C_4^2,C_4^3)\ra (-C^1_4,C^2_4,-C^3_4)\ .
\ee

Consistency of the theory therefore requires the target space theory
containing $C_4^r$ to be self-dual under $(C_4^1,C_4^2,C_4^3)\ra
(-C^1_4,C^2_4,-C^3_4)$.


The action \eq{l6} remains invariant under simultaneous $Sp(1)$
rotations of the 3-brane charge $\vec{Q}$ and the spacetime 4-form
potentials $\vec{C}_4$. In order to obtain a manifestly $Sp(1)$
covariant action we need to first eliminate the auxiliary fields. In
order to eliminate the auxiliary fields through their field
equations we observe the following useful properties of the $C$ and
$S$ functions:

\be
dC\se \e_{ij}V_i S_j \ ,\qquad dS_i\se \e_{ij}V_j C\ ,
\ee

where $d$ is the exterior derivative on $CP^1$ and $i,j$ its $SO(2)$ tangent
space indices, $V^i$ are the basis one-forms and $S=S_1+i S_2$.
The field equation then reads:

\be
Q_1 S_i-q_i z \se 0 \ .
\ee

Substituting back into the action \eq{l6} we find the following manifestly
$Sp(1)$ covariant form of the action:

\be
\cL\se |\vec{Q}|\sqrt{-\det{g}} - \ft1{4!}\e^{mnpq}
\vec{Q}\cdot\vec{\unC}_{mnpq}\ .
\ee

\subsection{$3$-brane in $D=5$ via vertical reduction from
$D=6$\label{vert}}

The construction of the $3$-brane in $D=5$ can proceed either by
directly solving the embedding and $\cF_1$ constraints obtained by
vertical reduction of the six-dimensional embedding constraint and
directly constructing the GS action from the reduced Wess-Zumino
term, or, equivalently, by vertical reduction of the six-dimensional
embedding matrix and Green-Schwarz action.


We begin with the first approach in which we vertically reduce the
six-dimensional embedding constraint. To this end, we first separate
off $E^{6}$ and identify

\be
\cF_1\;\;\equiv\;\;-\unE^6\ ,\qquad  H_2\;\;\equiv\;\; -T^{6}\ .
 \ee

The pull-back of the six-dimensional torsion identity $dE^6=T^6$
(since $\hat{\o}_{\hat{\una}}{}^6=0$) then automatically gives the
$\cF$ Bianchi identity \eq{fbi3}. Furthermore, the six-dimensional
embedding condition $\hat{E}_\a{}^{\hat{\una}}=0$ implies the
five-dimensional embedding and $\cF_1$-constraints:

\be
E_\a{}^{\una}\se 0\ ,\qquad \cF_{\a}\se 0\ .
\la{5dc}
\ee

Making the definition

\be
\hat{H}_5\;\;\equiv\;\;H_5+ E^{6}H_4 \ ,
\ee

it follows that the six-dimensional Wess-Zumino term
$\hat{W}_5=\hat{\unH}_5$ reduces to the $W_5$ given in \eq{p359} for
$t=1$. The five-dimensional constraints \eq{5dc} together with the
action formalism discussed in Section \ref{action}, provide a basis
for a self-contained formulation of the scalar multiplet of the
$3$-brane in five dimensions.

Without loss of generality we can parametrise the spinor-spinor part
of the embedding matrix as

\be
E_{\a}{}^{\ua} \se {1\over\sqrt{2}}\left[(u\otimes v_1)_{\a}{}^{\ua}
+(\c^5u\otimes v_2)_{\a}{}^{\ua}\right] +{1\over\sqrt{2}}h_{\a}{}^{\b}
\left[(\c^5 u\otimes v_1)_{\b}{}^{\ua} -(u\otimes v_2)_{\b}{}^{\ua}
\right]
\la{e5}
\ee

where $\ua$ is a five-dimensional symplectic Majorana index. In flat
target space, the dimension zero torsion equations \eqs{T}{h} yield

\bea
h_{\a\b}&=& A C_{\a\b} + iB (\c^{5})_{\a\b} + i h_{a}(\c^{a}\c^5)_{\a\b}
\ ,\qquad z=A+iB
\nn\w2
T_{\a\b}{}^a&=&-im^{a}{}_{b}(\c^{b})_{\a\b} + 2[(A\c^5+iB)\c^{ab}h_b]_{\a\b} \
,\nn\w2 m_{ab} &=& \eta_{ab} (1 + |z|^2 + h^2) -
2h_{a}h_{b}\ ,
\la{hTm}
\eea

where $h^2:=h^a h_a$. Just as in the six-dimensional case, the $v$'s
are not constrained by the torsion equations, and there exists an
extra $Sp(1)$ symmetry in the spinor-spinor part of the embedding
matrix enabling us to gauge away three of the five scalar
superfields in $(v_I{}^i,A,B)$ leaving two superfields related to
the two auxiliary fields of the scalar multiplet. For $h_{\a\b}$
given by \eq{hTm}, the infinitesimal local $Sp(1)$ transformations
take the form:

\bea
\d E_+ &=& \left(iN_1+(N_2+iN3)z\right)E_+-i(N_2-iN_3)\c^ah_aE_-\ ,
\nn\w2
\d v_I{}^j &=& iN^r(\s^r)_I{}^Jv_J{}^j\ ,
\nn\w2
\d z &=& (N_2-iN_3)(1+h^2)+2iN_1 z + (N_2+iN_3)z^2\ ,
\nn\w2
\d h_a &=& 2{\rm Re}\left[(N_2+iN_3)z\right]\,h_a\ ,
\la{sp1}
\eea

where $E_\pm=\ft12(1\pm\c^5)E$ and $N^r$ are three real superfields.
As we shall see below, \eq{sp1} is in agreement with the vertical
reduction of the M\"{o}bius symmetry found in the six-dimensional
case.

As discussed in Section \ref{emb}, the embedding constraint leaves the
worldvolume multiplet underconstrained and therefore we need to make
use of the $\cF_1$ constraint $\cF_\a=0$ subject to the appropriate
Bianchi identity $d\cF_1=\unH_2$ to obtain the off-shell scalar
multiplet. At dimension zero this reads

\be
T_{\a\b}{}^c \cF_{c} \se E_{\a}{}^{\ua}E_{\b}{}^{\ub} H_{\ua\ub}\ .
\ee

Tracing this equation with $\c^d$ (the $\c^{de}$ trace is
identically obeyed) yields

\be
m_a{}^b{\cF}_{b} \se -2 h_a\ ,
\la{ab3}
\ee

with $m_{ab}$ given in \eq{hTm}. Inverting we find

\be
\cF_a\se {-2h_a\over 1+|z|^2-h^2}\ .
\ee

Analysing also the dimension half components of the $\cF_1$ Bianchi
identity\footnote{A convenient trick \cite{rudy} is to replace this
problem by the equivalent problem of finding the conditions on the
torsion such that the only globally well-defined solution to
$d\O_2=0$ is $\O_2=0$.} one finds that the worldvolume multiplet is
an off-shell scalar multiplet whose components are two scalars and a
spinor, given by the leading components of the transverse
supercoordinates of the superembedding and the leading component of
the 0-form potential, and two auxiliary fields given by the leading
component of $z$.

We construct the action in the usual way. With $W^{(S)}_5$ given by
\eq{p359} for $t=1$, the dimension zero component of
$W^{(S)}_5=dK^{(S)}_4$ reads

\be
T_{\a\b}{}^d K^{(S)}\e_{dabc}\se \vec{Q}\cdot\vec{\unH}_{abc\a\b}-
3\cF_{[a}\vec{Q}\cdot\vec{\unH}_{bc]\a\b}\ ,
\la{w5}
\ee

where we have set $K^{(S)}_{abcd}= \e_{abcd}K^{(S)}$ and the
background is specified in \eq{hhh} and $\vec{Q}$ is the brane charge.
Tracing with $(\c^e)^{\a\b}$ we find

\be
K^{(S)}\se {Q_1(1-|z|^2+h^2)+q\bar{z}+\bar{q}z\over
1+|z|^2-h^2}\ .
\la{ks}
\ee

As mentioned before all the remaining components of $W=dK$
(including the higher dimensional ones) will then be identically
satisfied. To eliminate $h_a$ in favor of $\cF_a$ in \eq{ks}, we use
the following redefinition of the auxiliary fields:

\be
\hat{z}\se {z\over 2|z|^2}\left(\sqrt{(1
+h^2-|z|^2)^2+4|z|^2}-1-h^2+|z|^2)\right)\ .
\la{zhat}
\ee

Actually, as we shall see in the next section, $\hat{z}$ is
identical to the six-dimensional auxiliary field in the $v=1$ gauge
discussed earlier. Inserting \eq{zhat} in \eq{ks} we obtain the
kinetic term in Born-Infeld form with an auxiliary field dependent
prefactor as follows:

\be
K^{(S)} \se {Q_1(1-|\hat{z}|^2)+q\hat{\bar{z}}+\bar{q}\hat{z}\over
1+|\hat{z}|^2} \sqrt{-\det(\h_{ab}+\cF_a\cF_b)}\ ,
\ee

The corresponding GS Lagrangian is

\bea
\cL^{(S)} &=& {Q_1(1-|\hat{z}|^2)+q\hat{\bar{z}}+\bar{q}\hat{z}\over
1+|\hat{z}|^2}\sqrt{-\det g} \sqrt{\det(\d_a^b+\cF_a\cF^b)} \nn\w3
&&-{1\over 4!}\e^{mnpq}\vec{Q}\cdot(\vec{{\unC}}_{mnpq}+4\vec{{\unC}}_{mnp}
\cF_{q})\ ,
\la{slag}
\eea

where, as usual, $g$ is the standard induced metric on the bosonic
worldvolume and the 4-form potential is evaluated at $t=1$ in
\eq{ct}.


We now turn to the second approach to obtain the 3-brane action in
$D=5$ by vertical reduction of the 3-brane action in six-dimensions.
The key object object in this approach is the spinor-spinor part of
the embedding matrix. To study its vertical reduction, we start by
parametrising the element $\hat{u}_\a{}^{\hat{\ua}}$ of
$Spin(5,1)/Spin(3,1)$ in terms of an element $u_\a{}^{\ua}$ in
$Spin(4,1)/Spin(3,1)$ and a worldvolume superfield $K_a$
corresponding to the coset generator $\C^{a6}$:

\be
\hat{u}^{\ua}\se {1\over\sqrt{1+K^2}}(1+i\c^aK_a)u^{\ua}\ ,
\la{u6}
\ee

where the six-dimensional symplectic Majorana-Weyl index on the left
side has been identified with the five-dimensional symplectic
Majorana index on the right side. Since the vertical reduction does
not affect the $Sp(1)$ group elements, we have

\be
\hat{v}_I{}^j\se v_I{}^j\ .
\ee

We next write

\be
\hat{E}_\a{}^{\ua}\se M_\a{}^{\b}E_\b{}^{\ua}\ ,
\la{e6}
\ee

where $M_\a{}^\b$ is an invertible matrix which can be chosen such
that the five-dimensional $E_\a{}^{\ua}$ assumes the canonical form
given in \eq{e5}. Inserting the definitions made in \eq{u6} and the
six-dimensional embedding matrix $\hat{E}_\a{}^{\ua}$ given in
\eq{euv} (with $\hat{h}_{\a\b}$ given by \eq{hab}) into \eq{e6} one finds that

\bea
M &=& {1\over\sqrt{1+K^2}}\left(1-i(\hat{A}+
i\c^5\hat{B})\c^a\c^5K_a\right)\ ,
\nn\w2
h_{\a\b}&=& A+i\c^5 B + i\c^a\c^5 h_a\ ,
\la{mh}
\eea

where

\bea
K_a&=&{1-|\hat{z}|^2K^2\over 1+|\hat{z}|^2}\,h_a \;\;\equiv\;\; U h_a\ ,
\nn\w2
\hat{z}&=&{1-|\hat{z}|^2K^2 \over 1+K^2}\,z\quad\equiv\;\; V z\ .
\la{kz}
\eea

From these equations we obtain the solution:

\bea
U &=&{1\over 2h^2}\left(\D\pm(1+|z|^2-h^2)\right)\ ,
\nn\w2
V &=& {1\over 2|z|^2}\left(\D\pm(1+h^2-|z|^2)\right)\ ,
\nn\w2
\D&\equiv& \sqrt{(1+|z|^2 -h^2)^2+4h^2}\se
\sqrt{(1-|z|^2 +h^2)^2+4|z|^2}\ .
\la{uv}
\eea

It is also useful to note the characteristic equations:

\bea
1+h^2U^2&=&\D U\ ,\qquad 1-h^2U^2\se (1+|z|^2-h^2)U
\ ,\nn\w2
1+|z|^2V^2&=&\D V\ ,\qquad 1-|z|^2V^2\se (1+h^2-|z|^2)V\ .
\la{ch}
\eea

The local $Sp(1)$ transformations inherited from the six-dimensional
theory can now be shown to reproduce the infinitesimal version of
these transformations given in \eq{sp1}.

To reduce the vector-vector part of the embedding
matrix, we first write the vectorial counterpart of \eq{u6} as

\be
\hat{u}_{\hat{a}}{}^{\hat{\una}} \se
\L_{\hat{a}}{}^{\hat{b}}u_{\hat{b}}{}^{\hat{\una}} \ ,
\ee

where the index $\hat{a}\ra (a,5,6)$, and $\hat{u}$ belongs to
$SO(5,1)/SO(3,1)$ and $u$ to $SO(4,1)/SO(3,1)$ (embedded in $SO(5,1)$
such that $u_{\hat{a}}{}^{6}=\d_{\hat{a}}^6$). The basic spinor identity

\be
\hat{u}_\a{}^{\ua}\hat{u}_\b{}^{\ub}(\C^{\hat{\una}}C_6)_{\ua\ub}\se
(\C^{\hat{b}}C_6)_{\a\b}\hat{u}_{\hat{b}}{}^{\hat{\una}}
\la{tw}
\ee

and the Dirac matrices $\C^{\hat{a}}C_6\ra (\c^aC_4,C_4,i\c^5)$ implies that

\be
\L_{\hat{a}\hat{b}}\se{1\over 1+K^2}\mx{(}{ccc}{(1+K^2)\h_{ab}
-2K_aK_b&0&2K_b\w3 0 & 1+K^2 & 0 \w3
-2K_a & 0 & 1-K^2}{)}\ .
\ee

Assuming that the vector-vector part of the embedding matrix is given by
the canonical expression \eq{emp1} (in both five and six dimensions) it follows
that

\be
\hat{E}_a{}^{\una}\se \L_a{}^b E_b{}^{\una}\ ,\qquad \hat{E}_a{}^6\se \L_a{}^6\ .
\la{he}
\ee

Notice that $\L_{ab}$ is invertible but not orthogonal. Furthermore,
the basic embedding constraint imply that $\unE^{\una}
=\hat{E}^a\hat{E}_a{}^{\una}=E^a E_a{}^{\una}$, where $\hat{E}^a$
and $E^a$ are the induced co-vector frames on the worldvolume in the
six- and five-dimensional formulations, respectively. Comparing with
\eq{he} we find that

\be
E^a\se \hat{E}^b\L_b{}^a\ .
\la{ea}
\ee

Defining the components of $\cF=-\unE^{6}$ with respect to the frame
$E^a$ by

\be
\cF\;\;\equiv\;\; E^a\cF_a\ ,
\ee

and using \eq{he} and \eq{ea} we find that

\be
\cF_a\se -{2\over 1-K^2}K_a\ .
\ee

From \eq{kz} and \eq{ch} it follows that this equation is equivalent to
\eq{ab3}.

Thus, we conclude that the vertical reduction of the six-dimensional
Green-Schwarz action \eq{l6} is equivalent to the five-dimensional
action \eq{slag} found by direct construction. The appearance of the
Born-Infeld factor in the first term is a straightforward
consequence of the reduction of the induced metric (using the
relation given in \eq{gsm}):

\be
\hat{g}_{mn}\se \hat{E}_m{}^{\hat{\una}}\hat{E}_{n\hat{\unb}}
\se \hat{E}_m{}^a\hat{E}_n{}^b
\left(\hat{E}_a{}^{\una}\hat{E}_{b\una}+\hat{E}_a{}^6
\hat{E}_{b6}\right)
\se g_{mn}+\cF_m\cF_n\ ,
\ee

and the reduced Wess-Zumino term corresponds to the choice $t=1$ in
\eq{z12359}.

\subsection{Dualisation of the $3$-brane in $D=5$ and the linear
multiplet \label{dual}}

We shall now dualise the scalar potential $\cA_0$ and recover the
linear multiplet version of the theory. To this end, we eliminate the
auxiliary field $z$ using its equation of motion and obtain the simpler
Lagrangian

\be
\cL^{(S)} \se |\vec Q|\sqrt{-\det
g}\sqrt{\det(\d_a^b+\cF_a\cF^b)} -{1\over
4!}\e^{mnpq}\vec{Q}\cdot(\vec{\unC}{(1)}_{mnpq}+4\vec{\unC}_{mnp}\cF_{q})\ ,
\la{slag1}
\ee

where we recall the definition of $C^r_4{(1)}$ given in \eq{ct}. To
dualise $\cA_0$, we first relax the $\cF_1$ Bianchi identity and add a
Lagrange multiplier term as follows

\be
\cL^{(S)'} \se \cL^{(S)}-\cA_2(d\cF_1-\unH_2)\ .
\ee

Notice that under the background gauge transformation

\be
\d C_3\se d\L_2\ ,\qquad \d C_4\se -H_2\L_2\ ,\qquad \d\cF_1\se 0\
, \la{la2}
\ee

the variation of the scalar Lagrangian is
$\d\cL^{(S)}=-\underline{\L}_2(d\cF_1-\unH_2)$. Therefore the
Lagrangian $\cL^{(S)'}$ is invariant under \eq{la2} provided

\be
\d A_2\se -\underline{\L}_2\ \ .
\ee

$\cF_1$ can now be treated as an independent worldvolume field that can
be integrated out. This is achieved by using the algebraic field
equation for $\cF_1$ which can be put into the form

\be
\cF_a \se -{1\over |\vec{Q}|}\,\sqrt{1+\cF^2} \ \cH_a\ ,
\la{tr1}
\ee

where $\cH_1$ is the Hodge dual of the gauge-invariant three-form field
strength $\cF_3 = d\cA_2+ \vec{Q}\cdot\vec{\unC}_3$. Substituting this
result into the Lagrangian $\cL^{(S)'}$ and using $ C^r_4(1) =
C^r_4+C_1C^r_3$ we obtain the dualised Lagrangian

\be
\cL^{(S)'} \se \sqrt{-\det
g}\,\sqrt{\det(|\vec{Q}|^2\d_a{}^b-{\cH}_a{\cH}^b)}\  - \ {1\over
4!}\e^{mnpq}\vec{Q}\cdot(\vec{\unC}_{mnpq}+4\vec{\unC}_m\cF_{npq})\ ,
\la{linlag}
\ee

where the 4-form field strength in the Wess-Zumino term corresponds
to the value $t=0$ in \eq{ct}. The dualisation could also have been
obtained on-shell by using the relation \eq{tr1} to map the $\cF_1$
Bianchi identity into the $\cA_2$ field equation and the $\cA_0$
field equation into the $\cF_3$ Bianchi identity.

\subsection{Direct construction of the dual $3$-brane in
$D=5$\label{linear}}

The dualised Lagrangian $\cL^{(S)'}$ found in the previous section can
also be obtained directly from the superembedding formalism. To achieve
this we impose the standard embedding constraint $E_\a{}^{\una}=0$, but
we replace the $\cF_1$ constraint by the dual ${\cF}_3$ constraint
\eq{fc} subject to the appropriate Bianchi identity $d\cF_3=\unH_4$,
and obtain the off-shell linear multiplet. Thus, we start with the form
of the spinor-spinor part of the embedding matrix given in \eq{e5}, and
use the local $Sp(1)$ symmetry \eq{sp1} to set the $Sp(1)$ valued
superfields $v_I{}^i=\d_I{}^i$. At dimension $0$ the $\cF_3$ Bianchi
reads

\be
T_{\a\b}{}^c \cF_{abc} \se E_{\a}{}^{\ua}E_{\b}{}^{\ub}
E_a{}^{\una} E_b{}^{\unb}
\,\,\vec{Q}\cdot\vec{H}_{\underline{ab\a\b}}
\ee

where the background value of $H_4$ is given in \eq{hhh} and $\vec{Q}$
is the brane charge. This equation, being symmetric on the spinor
indices, can then be traced with either $\c^d$ or $\c^{de}$. This leads
to the following results:

\bea
\cF_{abc}&=& {2Q_1\over 1-|z|^2+h^2}\e_{abcd}h^d\ ,
\nn\w2
z&=& \ft12 (1-|z|^2+h^2)\,q\ ,
\la{ab2}
\eea

where $h_a$ and $z$ are defined in \eq{hTm}. Thus $z$ is no longer an
independent (auxiliary) field and $h_a$ is the non-linear dual of
the field strength of the 2-form potential. Thus the worldvolume
multiplet is a linear multiplet whose components are a scalar and
spinor given by the leading components of the transverse coordinates of
the superembedding and the leading component of the 2-form potential.

We construct the action following the recipe given in Section \ref{action}. The
appropriate $W^{(L)}_5$ is given by\eq{p359} for $t=1$. With
$W^{(L)}_5=dK^{(L)}_4$ we find that the only non-vanishing component of
$K^{(L)}_4$ is $K^{(L)}_{abcd}=\e_{abcd}K^{(L)}$, where

\be
K^{(L)} \se Q_1{1+|z|^2 -h^2\over 1-|z|^2+h^2}\ .
\ee

In terms of the Hodge dual ${\cH}_a$ of ${\cF}_{abc}$ defined by

\be
{\cF}_{abc} \se \e_{abcd}{\cH}^d\ ,
\la{hd}
\ee

the kinetic $K^{(L)}$ is given by the Born-Infeld style Lagrangian

\be
K^{(L)} \se \sqrt{-\det(|\vec{Q}|^2\h_{ab}-{\cH}_a{\cH}_b)}\ ,
\ee

and the total Green-Schwarz Lagrangian is given by \eq{linlag} found
by the dualisation procedure.

\section{5-brane in 7-dimensions\label{5in7}}

 In flat $D=7$ target space equation \eq{A}, which is the consequence of the
 dimension $0$  torsion Bianchi identity in the normal directions,
 implies that

\be
h_{\a\b}^{ij} \se \e^{ij}(\c^{abc})_{\a\b}h_{abc} +
u_{a}^{ij}(\c^{a})_{\a\b}\ .
\ee

The dimension $0$ torsion Bianchi identity in the tangential direction,
on the other hand, yields the result

\be
T_{\a i,\b j}{}^c \se -i\e_{ij} M^c{}_d\,(\c^d)_{\a\b}
-iM^{c,def}_{ij}\,(\c_{def})_{\a\b} \ ,
\ee

where

\bea
M_{ab} &=&\eta_{ab}-72k_{ab} +\ft12 \eta_{ab} u^2-(u\cdot u)_{ab}\ , \nn\w2
M_{ij}^{d,abc}&=& 12 h^{d[ab}\,u^{c]_+}_{ij}
+\eta^{d[a}\,u^b_{ik}\,u^{c]_+}{}_{\ell j}\,\e^{k\ell}\ ,
\eea

where $[abc]_+$ denotes the self-dual projection, and

\bea
k_a{}^b &:=&h_{acd}\,h^{bcd}\ , \nn\w2
u^2 &:=& u_a^{ij}u^a_{ij}\ ,\quad \quad (u\cdot u)_{ab}\;\; :=\;\;
u_a^{ij}u_{bij}\ .
\eea

One can show that the analysis of the dimension $1/2$ and higher
torsion Bianchi identities do not put the system on shell, and that the
superfields $h_{abc}$ and $u_a^{ij}$ remain underconstrained. In
\cite{ch}, the superfield $u_a^{ij}$ was set equal to zero by hand, and
consequently, the on-shell system describing the $(1,0)$ tensor
multiplet on the worldvolume was obtained. This procedure is not
altogether satisfactory from a geometrical point of view. Here we will
show that the introduction of super $3$-form field strength ${\cF}_3$
defined in \eq{fbi3} subject to the constraint \eq{fc} indeed leads to
the equation $u_a^{ij}=0$. As discussed earlier, ${\cF}_3$ is a natural
geometrical ingredient for constructing the necessary Wess-Zumino term
and the constraint it satisfies can be understood from geometrical
considerations involving a superstring ending on the superfivebrane.

At dimension $0$, the ${\cF}$-Bianchi identity \eq{dfh} implies that

\bea
M_a{}^d\;{\cF}_{bcd} &=& 24 h_{abc}\ ,
\la{5a}\w2
M_{ij}^{f,abc}\,{\cF}_{def} &=& -2 u_{ij}^{[a}\,\d^{bc]_+}_{de}\ .
\la{5b}
\eea

Some useful definitions and relations are

\bea
&& (\c_{abc})_{\a\b}\se -\ft16\,\e_{abcdef}\,(\c^{def})_{\a\b}\ ,
\quad\quad h_{abc}\se \ft16\,\e_{abcdef}\,h^{def}\ ,
\nn\w3
&&k_a{}^d\,h_{bcd}\se k_{[a}{}^d\,h_{bc]d}\se -\ft16\,\e_{abc}{}^{def}\,
k_{d}{}^g\,h_{efg}\ ,
\nn\w3
&& h_{abe}\,h^{cde} \se \d_{[a}^{[c}\,k_{b]}{}^{d]}\ ,
\quad\quad
k_{ac}\,k^{bc} \se \ft16 \d_a^b\,{\rm tr}\,k^2\ ,\qquad
\eta^{ab} k_{ab}\se 0\ .
\la{lemmas}
\eea

We shall now show that $u^a_{ij}$ vanishes as a consequence of \eq{5a}
and \eq{5b}. We begin by rewriting \eq{5a} (assuming that $M_{ab}$ is
invertible) as

\be
{\cF}_{abc}  \se  24(M^{-1})_a{}^d\,h_{bcd}\ .
\la{a0}
\ee

Multiplying this equation with suitable $M$ matrices twice yields

\be
M_{(a}{}^d\,h_{b)cd}  \se  0\ .
\la{a1}
\ee

Noting that $m_{(a}{}^d\,h_{b)cd}=0$, which can be deduced from the
lemmas listed in \eq{lemmas}, we find from \eq{a1} that

\be
(u^2)_{(a}{}^d\,h_{b)cd}  \se 0\ .
\la{a2}
\ee

Next, we  turn to the analysis of \eq{5b}. Using \eq{a0} in \eq{5b} and
multiplying with a suitable $M$, and symmetrizing with respect to a
pair of indices in which the left hand side is manifestly antisymmetric
and therefore drops out, we arrive at

\be
u^{[a}_{ij} M_{(d}{}^b \d_{e)}^{c]_+} \se 0\ .
\la{a10}
\ee

Using equations \eq{a2} and \eq{a10} one can show, after some algebra, that

\bea
(u^2)_{ab}&= &A k_{ab} + B\h_{ab} \la{u}\w2
M_{ab}&=&(1+2B)\h_{ab} -(A+72) k_{ab}\ ,
\eea

where $A=\tr (u^2 k)/\tr (k^2)$ and $B=\tr (u^2)/6$. Furthermore,
either $A$ and $B$ are both zero, in which case $u^a_{ij}$ itself
must be zero, or $A=-72$ and $B$ is undetermined. However, a further
property of the matrix $(u^2)_{ab}$ that it has maximal rank three
as a consequence of its definition. This implies that

\be
(u^2)_{[a_1}{}^{[b_1}\cdots (u^2)_{a_4]}{}^{b_4]} \se 0\ .
\ee

This equation can be used to show that, if $A=-72$, then $B= \pm
A\sqrt{\tr(k^2)/6}$. This implies that $(u^2)_{ab}$ is proportional
to a projection operator,

\be
(u^2)_{ab} \se \pm A\sqrt{\tr(k^2)/6}(1\pm \bar k)\ ,
\la{kb}
\ee

where $\sqrt{\tr(k^2)/6}\bar k=k$. The matrix $\bar k$ therefore square
to the identity and is traceless. Bringing it to a canonical form, one
can then show the positivity properties of $(u^2)_{ab}$ rule out the
solution \eq{kb} so that, finally

\be
u^a_{ij} \se 0\ .
\ee

Using this result in the dimension $1/2$ and $1$ torsion Bianchi
identities, one finds the following field equations:

\bea && m^{ab} (\c_b)^{\a\b}\,\L_{bj} \se 0\ ,
\nn\w2
&& m^{ac}\,m_c{}^b\;
X_{ab}{}^{\perp} \se \ft{i}4\,m^{ab}\,(\c_a)^{\a\b}\,U_{b,\a\b}\ ,
\nn\w2
&& m^{ab} \,\nabla_a h_{bcd} \se
-\ft{i}{48}\,m^{ab}\left(\c_{cd}\c_b\right)^{\a\b}\;U_{a,\a\b}\ ,
\la{feqs}
\eea

where $X_{a,b}{}^{\perp}$ was defined in \eq{defx}, and where

\bea
m_a{}^b &=& \d_a^b-72 h_{acd} h^{bcd}\ , \nn\w2
U_{a,\a\b} &=&
\left[h_{cde}(\c^{cde}\c^b)_\a{}^\c\,
\L_{a\c i}\L_{b\b j}\,\e^{ij}+ \ft12(\a\ \leftrightarrow \b)\right] \ .
\eea

These equations are equivalent to those of \cite{ch} up to field
 redefinitions.

 One can also construct an action for an {\it unconstrained} 2-form potential
 $\cA_2$ such that its field equation is equivalent to the
 Bianchi identity $d\cF_3=\unH_4$ upon the imposition of a non-linear
 self-duality condition. Since the 4-form and 7-form field strengths
 in $D=7$ and $D=11$ obey formally equivalent Cartan integrable systems,
 the result found \cite{cns} for the 5-brane in $D=11$ carries over
essentially unchanged to the 5-brane in $D=7$ (see also \cite{ss2}
for a review). The resulting for the 5-brane in $D=7$ is

 \be
 \cL\se {1\over 2}\sqrt{-\det{g}}\cK - {1\over 6!}\e^{mnpqrs}\left(
 \unC_{mnpqrs}+ 10\unC_{mnp}\cF_{qrs}\right)\ ,
 \la{s5}
 \ee

 where the kinetic term is given by

 \be
 \cK\se\sqrt{1+\ft1{12}\cF^2+\ft1{288}(\cF^2)^2
 -\ft1{96}\cF_{abc}\cF^{bcd}\cF_{def}\cF^{efa}}\ ,
 \la{ck}
 \ee

and the Wess-Zumino term corresponds to the Wess-Zumino form $W_7$
given in \eq{p359}. Varying the unconstrained 2-form potential
$\cA_2$ one finds the second order tensor field equation

\be
d\left(\star {\del\cK\over \del \cF_3}\right)\se
\unH_4\ .
\la{te}
\ee

The self-duality condition
interchanging this field equation and the $\cF_3$ Bianchi identity
therefore must read

\be
\star \cF_3\se  {\del\cK\over \del \cF_3}\ .
\la{scsd}
\ee

For the particular form of $\cK$ in \eq{ck} the self-duality
condition is consistent (in the sense that $\star^2=1$). In fact,
\eq{scsd} is equivalent to \eq{a0}, and one can show that the second
order tensor field equation \eq{te} and the $z^{\unM}$ field
equations following from \eq{s5} are equivalent to the field
equations \eq{feqs} obtained from the superembedding upon imposition
of \eq{scsd}. The $\kappa$ transformation $\d z^{\unM}=\k^{\unM}$
and $\d \cA_3= i_\k \unC_3$ vanishes provided \eq{scsd} holds. The
$\k$ invariance can be understood from the embedding formalism by
solving for $K_6$ from \eq{w}. One then finds that the only
non-vanishing component is $K_{abcdef}=\e_{abcdef}K$ with

\be
K\se \sqrt{1+\ft1{24}\cF^{abc}\cF_{abc}}\ .
\la{k6b}
\ee

This expression can be shown to equal $\cK$ given in \eq{ck} upon
imposition of \eq{scsd}. The action formula \eq{9b} cannot be
applied, however, to derive the tensor field equations, since the
multiplet is on-shell {\it and} since there are on-shell constraints
on the tensor field strength at dimension zero which affect the
functional form of $K_6$.

\section{9-brane in 11-dimensions\label{9in11}}

We begin with the analysis of \eq{tors0} which yields

\bea
h_{\a\b}&=&h_{abc}(\c^{abc})_{\a\b}\ ,
\nn\w2
T_{\a\b}{}^a &=& m^a{}_b(\c^b)_{\a\b}+m^a{}_{b_1\cdots b_5}
(\c^{b_1\cdots b_5})_{\a\b}\ ,
\nn\w2
m_{ab}&=&(1+6h^2)\eta_{ab}-12k_{ab}\ ,
\nn\w2
m_{a, b_1\cdots b_5}&=& 6h_{a[b_1 b_2}\,h_{b_3 b_4 b_5]_+}
-9\eta_{a[b_1}h_{b_2 b_3 c}\,h^{c}{}_{b_4 b_5]_+}\ ,
\eea

where  $[abcdef]_+$ denotes the self-dual projection,
$h^2 := h_{abc} h^{abc}$ and

\be
k_{ab} \;\;:=\;\; h_{acd}\,h^{bcd}\ .
\la{k2}
\ee

At this stage we have an underconstrained system described by a
single unconstrained Goldstone superfield in $N=1$, $d=10$
superspace.

Let us begin by assuming that the 9-brane intersects an ordinary open
membrane in $D=11$. As explained in section {open}, this implies the
${\cF}$-Bianchi identity \eq{dfh} subject to the constraint \eq{fc}. At
dimension zero this leads to the equations

\bea
&&m_a{}^d \cF_{dbc}\se -12h_{abc}\ ,
\la{6a}\w2
&& m^{c,b_1 \cdots b_5}\cF_{cab}\se 2h^{[b_1 b_2 b_3}\d^{b_4 b_5]_+}_{a\,b}\ .
\la{6b}
\eea

The manipulation of \eq{6a}, in the same fashion as described in the
previous section in the analysis of \eq{5a}, gives

\be
k_{(a}{}^d\,h_{b)cd} \se 0\ .
\la{b1}
\ee

On the other hand, substituting the expression for ${\cF}$ obtained
from \eq{6a} into \eq{6b}, and multiplying once with a suitable
$m$-matrix and  symmetrizing in a pair of indices such that the left
hand sides drops out, we find

\be
h^{[abc}\,m_{(f}{}^d\,\d_{g)}^{e]_+} \se 0\ .
\la{b2}
\ee

Multiplying with $h_{abc}$ and then tracing two of the remaining free
indices gives, in matrix notation for $k_{ab}$,

\be
k^2 -({\rm tr}\,k)\,k \se 0\ ,
\la{b3}
\ee

which implies that the matrix $k$ has rank 1 and therefore can be
written in the form

\be
k_{ab} \se u_a u_b\
\la{k9}
\ee

for some vector $u^a$. Using this in \eq{b1} then leads to the result

\be
h_{abc} \se 0\ .
\la{h9}
\ee

This result can be shown to freeze the Goldstone superfield thereby
leading to only global degrees of freedom. To see this, we proceed
by analysing the constraints at the linearised level. We begin by
writing \eq{np} as

\be
D_\a \Phi \se \Psi_\a\ ,
\la{np2}
\ee

where we have defined $\Psi_\a := i(\C^{\perp})_{\a\b'}\Th^{\b'}$.
The linearised form of \eq{emp2}, on the other hand, gives

\be
h_{\a\b} \se D_{[\a} \Psi_{\b]}\ .
\ee

Therefore, using $h_{\a\b}=0$, together with \eq{np2} and the
superalgebra obeyed by the supercovariant derivatives, we find

\be
D_\a D_\b \Phi \se \ft{i}2 (\c^a)_{\a\b}\,\del_a\Phi\ .
\la{dec1}
\ee

Applying a third spinorial derivative and making repeated use of
\eq{np2} and \eq{dec1} yields

\bea
D_{\d}D_{\b}D_{\a} \F & = & \frac{i}{2} (\C^{a}) _{\b\a}\,
\del_{a}\Psi_{\d}
\nn\\
&=& -D_{\b}D_{\d}D_{\a}\Phi + i (\c^{a})_{\d\b}
                                           \,\del_{a}\Psi_{\a}
\nn\\
&=& - \frac{i}{2} (\c^{a})_{\d\a}\del_{a}\,\Psi_{\b}
        + i (\c^{a})_{\d\b}\, \del_{a}\Psi_{\a}\ .
\la{3d}
\eea

Comparing the first and the third lines one obtains

\be (\c^{a})_{\d\b}\,\del_{a}\Psi_{\a} \se \ft12
(\c^{a})_{\a\b}\,\del_{a}\Psi_{\d} + \ft12(\c^{a})_{\a\d}\,\del_{a}\Psi_{\b}\ .
\ee

Contracting the last equation with $(\c_{c})^{\c\d}$ and then with
$\d^{\b}_{\c}$ gives

\be
15 \,\del_{c}\Psi_{\a}\se
(\c_{c}{}^{d})^{\d}{}_{\a}\,\del_{d}\Psi_{\d}
\la{dec2}\ .
\ee

The contraction of \eq{dec2} with $(\c^{c})^{\b\a}$ implies the Dirac equation

\be
(\c^{c})^{\b\a}\del_{c}\Psi_{\a}\se 0\ ,
\ee

while contracting with  $(\C_{d})^{\b\a}$ gives

\be (\C_{d})^{\b\a}\del_{c}\Psi_{\a} +
(\c_{c})^{\b\a}\del_{d}\Psi_{\a} \se 0\ .
\ee

Further contraction of the last equation with $(\c_{a})_{\d\b}$ implies

\be
\h_{ad}\partial_{c}\Psi_{\d} +
(\C_{ad})_{\d}{}^{\a}\partial_{c}\Psi_{\a} + \h_{ac}\partial_{d}\Psi_{\d}
+ (\C_{ac})_{\d}{}^{\a}\partial_{d}\Psi_{\a} \se 0\ .
\ee

Contracting the last equation with $\h^{ad}$  and using the Dirac
equation we find

\be
\del_{c}\Psi_{\d} \se 0\ .
\la{f0}
\ee

Thus $\Psi$ must be a constant showing that the fermionic degrees of
freedom of the brane are frozen. To demonstrate a similar result for
the scalar we start from

\be
D_{\c}D_{\d}D_{\b}D_{\a}\F\se 0\
\ee

which follows from \eq{3d} and \eq{f0}. Symmetrizing in $(\c\d )$
and $(\b\a )$ and using the supersymmetry algebra of the
supercovariant derivatives we obtain

\be
(\c^{a})_{\c\d}(\c^{b})_{\b\a}\del_{a}\del_{b}\Phi \se 0\ ,
\ee

which in turn implies

\be
\partial_{a}\partial_{b}\F\se 0\ .
\ee

This represents a flat plane moving with a constant velocity. Notice
that it is a non-trivial fact that the brane allows for a zero-mode
spacetime momentum.


\section{Conclusions\label{concl}}


In this paper we have shown systematically how the embedding
constraint together with a constrained field strength $\cF$ in the
worldvolume provide a full description of the worldvolume multiplet
for branes with codimension one for which the standard embedding
constraint generically leads to an underconstrained system. The
constrained field strength arises in the context of open membranes
(or particles for $p=3$) ending on the codimension one branes.

For the 3-brane in five dimensions we have seen that there are two
possible choices for $\cF$, a 1-form or a 3-form. The 3-form option
gives rise to a worldvolume $N=1,d=4$ linear multiplet which is
off-shell; this is therefore the L-brane with 4 worldvolume
supersymmetries. The dualisation of the linear multiplet gives an
worldvolume $N=1,d=4$ scalar multiplet based on the 1-form option.
This multiplet is off-shell in contrast to the L-branes with 8
supersymmetries where dualisation of the antisymmetric tensor gauge
field to a scalar leads to an on-shell hypermultiplet \cite{hrs}. We
have also shown that the scalar version of the 3-brane in five
dimensions can be obtained by vertical reduction of a 3-brane in six
dimensions. It is worth noting that, from this perspective, the
five-dimensional $\cF_1$ constraint emerges as a component of the
standard embedding condition for a six-dimensional target.

The 3-brane completely breaks the $Sp(1)_R$ symmetry of the $N=1$,
$D=5$ supersymmetry algebra (the same situation arises for the
3-brane in $N=(1,0)$, $D=6$). The resulting three Goldstone scalars
parametrise the group element $v$ appearing in the spinor-spinor
part of the embedding matrix given in eq. \eq{e5}. The local $Sp(1)$
transformation $v\ra g v$, however, turns out to be an invariance of
the embedding, as we saw by writing the embedding matrix on the
manifestly $Sp(1)$ invariant form \eq{sp1inv}. In the scalar case we
found that the auxiliary fields parametrise a two-sphere. Upon
elimination of the auxiliary fields we then found an $Sp(1)$
covariant Green-Schwarz action, which we dualised to obtain the
linear multiplet formulation.

For the 5-brane in seven dimensions we have shown that the
$\cF_3$-constraint restricts the unconstrained scalar superfield
determined by the embedding constraint to be an on-shell $N=(1,0),
d=6$ tensor multiplet. The resulting equations of motion are in
agreement with those derived earlier in \cite{ch} where they were
obtained by imposing an additional torsion constraint by hand. We
also have constructed the 5-brane action analogous to the one given
for the 5-brane in $D=11$ \cite{cns}.

Finally, we have seen that the $\cF_3$-constraint for a 9-brane in
eleven dimensions severely restricts the worldvolume multiplet in
such a way that the degrees of freedom are frozen out such that the
only remaining degree of freedom is a spacetime momentum.

This suggests a connection with the Horava-Witten picture of a
9-brane as a boundary of the $D=11$ spacetime \cite{hw}. However,
the vector multiplets which arise on the boundary need to be
included in the superembedding formalism. To this end, it would be
interesting to investigate the consequences of modifying the $\cF_3$
constraint by including an $g_{YM}^{1/2}\tr(F^2)$ term to its
Bianchi identity.

Finally, we note that there is another kind of 9-brane, known as the
M9-brane, which we have not considered in this paper. This is a
wrapped domain wall solution of massive $D=11$ supergravity in the
sense of \cite{blo}, and it is closely related to an $M2$-brane in
which a $U(1) $ isometry direction in the target space is gauged. It
would be useful to find the fully nonlinear and supersymmetric
action for this brane, which is still lacking, within the framework
of the superembedding formalism.

\pagebreak


\section*{Appendix A}


In this appendix we give our conventions for the Dirac matrices that
are relevant to co-dimension one embeddings studied in this paper.
We also describe the splitting of the target space spinor space into
the tangential and normal directions appropriate to these
embeddings.

\subsection*{${\bf D=3 \ra d=2}$}

A 2 component Majorana spinor $\psi$ in $D=3$ splits as

\be
\psi_{\ua} \ \ \ra \ \
\left(\ba{c}
\psi_+ \\ \psi_-
\ea\right)
\ee

where $+$ labels the tangent direction and $-$ labels the normal
direction in the superembedding. The $D=3$ gamma matrices $\C^a C$
split as

\bea
&& \C^a_{++}= \d^a_+\ , \quad\quad \C^a_{--}= \d^a_-\ , \quad\quad
   \C^a_{+-}= \C^a_{-+}=0 \ ,  \quad a=0,1\ ,
\nn\w4
&& \C^3_{++}= \C^3_{--}=0 \ , \quad\quad \C^3_{+-}= \C^3_{-+}=1 \ .
\eea

We also define $\d^a_{+\!\!\!\!_{+}}=\ft12(1,1)$

\subsection*{${\bf D=4\ra d=3}$}

A $4$-component Majorana spinor $\psi_{\ua}$ in $D=4$ splits as

\be
\psi_{\ua} \ \ \ra \ \
\left(\ba{c}
\psi_\a \\ \psi_{\a'}
\ea\right)
\ee

where $\a$ labels a $2$-component Majorana spinor in the tangent
direction and $\a'$ labels another $2$-component Majorana spinor, this
one being in the normal direction. The latter spinor is in a
representation of the worldvolume Lorentz group which is equivalent to
that carried by the first spinor. The $\C$-matrices split as

\be
(\C^a)_{\ua}{}^{\ub}\se \left(\ba{cc} \c^a & 0\\
                              0          & \c^a \ea \right)
\quad\quad
(\C^4)_{\ua}{}^{\ub} \se \left(\ba{cc} 0 &  I\\
                                    I &  0\ea \right)
\quad\quad
C_{\ua\ub} \se \left(\ba{cc} C_3 & 0\\
                              0    & C_3\ea \right)
\ee

where $C_3$ is the standard charge conjugation matrix in three
dimensions and $\C^{\una}C_4$ are symmetric.

\subsection*{${\bf D=6 \ra d=4}$}

As discussed in detail in Section \ref{3in6}, the splitting of an
$Sp(1)$ symplectic Majorana-Weyl spinor $\Psi_i$ in $D=6$ involves the
$R$-symmetry doublet index $i$. Such a spinor consist of two Dirac
spinors obeying

\bea
\Psi_{c\,6}^i &=& -\e^{ij}\Psi_j\ ,
\nn\w2
\C^7\Psi_i&=&\Psi_i\ ,
\la{smw}
\eea

where $\Psi^i_{c6}=C_6\bar{\Psi}^{i\, T}$ is the charge-conjugated
spinor, $\C^{\underline{abcdef}}=\e^{\underline{abcdef}}\C^7$ and
$\c^{abcd}=i\e^{abcd}$ with $\e^{012356}=\e^{0123}=1$. The Dirac
matrices $\C^{\una}$ in $D=6$ can be written in terms of the $d=4$
Dirac matrices as

\bea
\C^a &=& \mx{(}{ll}{0 & \c^a\c^5 \\ -\c^a\c^5 & 0}{)}
\ ,\qquad
\C^5\se \mx{(}{ll}{0 & 1 \\ 1 & 0}{)}
\ ,\qquad
\C^6\se \mx{(}{ll}{  0 & i\c^5 \\ -i\c^5 & 0}{)}
\ ,\nn\w2
C_6 &=& \mx{(}{ll}{ 0 & -C_4\\ C_4 & 0}{)}\ ,
\eea

where $a=0,1,2,3$ and $C_6$ is symmetric and $\C^{\una}C_6$ are
anti-symmetric. The symplectic Majorana-Weyl condition \eq{smw}
translates into the following reality condition on the Majorana
conjugate spinors in $d=4$:

\be
\Psi^i_{c\,4}\se \c^5\e^{ij}\Psi_j\ .
\la{m4}
\ee

Since \eq{m4} implies that $\big(\c^5\Psi^i\Big)_{c4}=-\e^{ij}\Psi_j$
we can define two $d=4$ Majorana spinors as follows:

\bea
\Psi_{+}\;\equiv\; P_+^i\Psi_i \se
{1\/\sqrt2}\left(\Psi_1+\c^5\Psi_2\right)\ ,
\nn\w2
\Psi_{-}\;\equiv\; P_-^i\Psi_i \se
{1\/\sqrt2}\left(-\Psi_2+\c^5\Psi_1\right)\ .
\la{m5}
\eea

Note that in the cases of $D=5,6,7$ where the target space spinors are
$Sp(1)$ symplectic Majorana spinors, we shall use a notation where
$\ua$ denotes a composite Lorentz spinor and $Sp(1)$ doublet index.
Thus, for instance the symmetric composite Dirac matrices
$(\C^{\una}C)_{\ua\ub}$ splits into an anti-symmetric Dirac matrix
$(\C^{\una}C)_{\ua\ub}$ times $\e_{ij}$.

\subsection*{${\bf D=5 \ra d=4}$}

An $Sp(1)$ Majorana-Weyl spinor in $D=5$ reduces to an $Sp(1)$
symplectic Majorana spinor $\Psi_i$ in $D=5$ and splits essentially in
the same way as described in \eq{m5}. This expression is exactly
applicable if we choose the anti-symmetric charge conjugation matrix in
$D=5$ to be

\be
C_5 \se \c^5 C_4\ .
\ee

The five-dimensional symplectic Majorana condition then reads:

\be
\Psi^i_{c5}\se \e^{ij}\Psi_j\ .
\ee

We take the $D=5$ Dirac matrices (without the $Sp(1)$ doublet indices)
to be $\C^{\una}=(\c^a,\c^5)$. Note that $\C^{\una}C_5$ are symmetric.

\subsection*{${\bf D=7 \ra d=6}$}

An $Sp(1)$ symplectic Majorana spinor in $D=7$ decomposes into two
independent $Sp(1)$ symplectic Majorana-Weyl spinors of opposite
chiralities in $d=6$. Using chiral notation in which the spinor indices
which indicate the chiralities and cannot be raised or lowered, we
write

\be
\psi_{\ua} \ \ \ra \ \
\left(\ba{c}
\psi_\a \\ \psi_{\a'}
\ea\right)\equiv \left(\ba{c}
\l_{\a i} \\ \l^\a_i
\ea\right)\ .
\ee

where the upper and lower spinor indices $\a=1,...,4$ represent
left- and right-handedness. The spinors $\l_{\a i}$ and $\l^\a_i$ are
independent of each other. Using the chirally projected $\c$-matrices in $d=6$,
we can choose the $D=7$ $\C$-matrices as

\bea
(\C^a)_{\ua}{}^{\ub}&=&\d_i{}^j\mx{(}{cc}{ 0&(\c^a)_{\a\b}\\
(\c^a)^{\a\b}&0}{)}\ ,
\nn\w2
(\C^7)_{\ua}{}^{\ub}&=&\d_i{}^j\mx{(}{cc}{\d_\a^\b&0\\
0&-\d^\a_\b}{)}\ .
\eea

The anti-symmetric charge-conjugation matrix is

\be
C_{\ua\ub} \se \e_{ij}\mx{(}{cc}{ 0& \d_\a^\b\\\d^\a_\b &0}{)}\ .
\ee

Note that the Dirac $\C^{\una}C_7$ (without the $Sp(1)$ doublet index)
are anti-symmetric. Spinor indices in $D=7$ are raised
and lowered by the charge conjugation matrix so that

\bea
(\C^a)_{\ua\ub}&=&\e_{ij}\mx{(}{cc}{ (\c^a)_{\a\b}&0\\
0&(\c^a)^{\a\b}}{)}\ ,
\nn\w2
(\C^7)_{\ua\ub}&=&\e_{ij}\mx{(}{cc}{0&\d_\a^\b\\
-\d^\a_\b&0}{)}\ ,
\eea

\subsection*{${\bf D=11 \ra d=10}$}

A Majorana spinor in $D=11$ decomposes into two independent
Majorana-Weyl spinors of opposite chiralities in $d=10$. Using chiral
notation, we write

\be
\psi_{\ua} \ \ \ra \ \
\left(\ba{c}
\psi_\a \\ \psi_{\a'}
\ea\right)\equiv \left(\ba{c}
\l_{\a} \\ \l^{\a}
\ea\right)\ ,
\ee

where the upper and lower spinor indices $\a=1,...,16$ represent
left- and right-handedness. The spinors $\l_{\a}$ and $\l^{\a}$ are
independent of each other. Using the chirally projected $\c$-matrices in $d=10$,
we can choose the $D=11$ $\C$-matrices as

\bea
(\C^a)_{\ua}{}^{\ub}&=& \mx{(}{cc}{ 0&(\c^a)_{\a\b}\\(\c^a)^{\a\b}&0}{)}\ ,
\nn\w2
(\C^{11})_{\ua}{}^{\ub}&=& \mx{(}{cc}{\d_\a^\b&0\\0&-\d^\a_\b}{)}\ .
\eea

The charge-conjugation matrix is

\be
C_{\ua\ub}=\mx{(}{cc}{ 0& \d_\a^\b\\-\d^\a_\b &0}{)}
\ee

and $\C^{\una}C$ are symmetric and given by

\bea
(\C^a)_{\ua\ub}&=&\mx{(}{cc}{ -(\c^a)_{\a\b}&0\\0&(\c^a)^{\a\b}}{)}\ ,
\nn\w2
(\C^7)_{\ua\ub}&=&\mx{(}{cc}{0&\d_\a^\b\\ \d^\a_\b&0}{)}\ .
\eea


\section*{Appendix B}


In this appendix we give the results of solving the torsion equations
\eq{tors}. We first split these equations into tangential and normal
components by contracting with $(E^{-1})_{\unC}{}^{C}$ and
$(E^{-1})_{\unC}{}^{C'}$, respectively. The result is:

\bigskip
{\bf Tangential Projections:}

\begin{itemize}

\item {\it dimension $0$:}

\be
T_{\a\b}{}^{c} \se -i\left\{
(\C^{c})_{\a\b}-h_{\a\c}(\C^c)^{\c\d}h_{\d\b}\right\}
\la{T1}
\ee

\item {\it dimension $\frac12$:}

\bea
T_{\a\b}{}^{\c}&=& -ih_{\d(\a}(\C^a)^{\d\c}\L_{a\b)} +Y_{\a\b}{}^{\c}+
h_{\d(\a}(\C^a)^{\d\c}Y_{a\b)}{}^{\perp}\ ,
\w2
T_{a\b}{}^{c} &=& -i\L_{a\c}(\C^ch)^{\c}{}_{\b}+Y_{a\b}{}^{c}\ .
\la{T2}
\eea

\item {\it dimension $1$:}

\bea
T_{a\b}{}^{\c}&=& \frac{1}{2}\left\{
h_{\b\d}(\C^b)^{\d\c}X_{a,b}{}^{\perp}+i\L_{a\d}(\C^b)^{\d\c}\L_{b\b}\right\}
+Y_{a\b}{}^{\c}-\frac{1}{2}\L_{a\d}(\C^b)^{\d\c}Y_{b\b}{}^{\perp}
\ ,\w2
T_{ab}{}^{c}&=& -i\L_{a\a}(\C^c)^{\a\b}\L_{b\b}+Y_{ab}{}^{c}\ .
\la{T3}
\eea

\item {\it dimension $\frac32$:}

\be
T_{ab}{}^{\c} \se -\L_{[a\d}X_{b],c}{}^{\perp}(\C^c)^{\d\c}+
Y_{ab}{}^{\c}\ .
\la{T4}
\ee

\end{itemize}

\bigskip

{\bf Normal Projections:}

\begin{itemize}

\item{\it  dimension $0$:}

\be
h_{(\a\b)}\se 0\ ,
\la{A}
\ee

\item {\it dimension $\ft12$:}

\bea
\nab_{(\a}h_{\b)\c} &=& \frac{1}{2} T_{\c(\a}{}^{a}\L_{a\b)}
-\frac{1}{2} T_{\a\b}{}^{a}\L_{a\c} + Z_{\a\b ,\c}\ ,
\la{B}\w2
X_{\a,b}{}^{\perp} &=& i\L_{b\a}-Y_{b\a}{}^{\perp}\ .
\la{D}
\eea

\item {\it dimension $1$:}

\bea
\nab_{\a}\L_{a\b} &=& \nab_{a} h_{\a\b}+\frac{i}{2}
T_{\a\b}{}^{b} X_{a,b}{}^{\perp}
\nn\w2
&& -i \left\{\L_{a\c}(\C^{b})^{\c\d}h_{\d\a}\L_{b\b}
+\frac12\L_{a\c}(\C^{b})^{\c\d}h_{\d\b}\L_{b\a} \right\}
+Z_{a,\a\b}\ ,\la{C}\w2
X_{[a,b]}{}^{\perp} &=& \frac{1}{2} Y_{ab}{}^{
D}\ .
\la{E}
\eea

\item {\it dimension $\frac32$:}

\be
\nab_{[a}\L_{b]\a}\se\frac{1}{2}\left\{
\L_{[a\b}X_{b],c}{}^{\perp}(\C^c)^{\b\c}h_{\c\a}
+i \L_{a\b}(\C^c)^{\b\c}\L_{b\c}\L_{c\a}\right\} +Z_{ab,\a}\ .
\la{F}
\ee

\end{itemize}

where we have collected the contributions from curved target space
backgrounds in the quantities $Y_{AB}{}^{C}$ and $Y_{AB}{}^{C'}$ and
$Z_{AB,C}$ (which vanish in flat target space) defined by

\bea
Y_{AB}{}^C &\equiv& (-1)^{A(B+\unB)}E_{B}{}^{\unB}E_{A}{}^{\unA}
T_{\unA\unB}{}^{\unC}(u^{-1})_{\unC}{}^C -(-1)^{A(B+1)}
E_{B}{}^{\ub}E_{A}{}^{\ua}T_{\ua\ub}{}^{\unc}(u^{-1})_{\unc}{}^C
\nn\w3
Y_{AB}{}^{C'} &\equiv&
(-1)^{A(B+\unB)}E_{B}{}^{\unB}E_{A}{}^{\unA}
T_{\unA\unB}{}^{\unC}(u^{-1})_{\unC}{}^{C'} -(-1)^{A(B+1)}
E_{B}{}^{\ub}E_{A}{}^{\ua}T_{\ua\ub}{}^{\unc}(u^{-1})_{\unc}{}^{C'}\ .
\la{1}
\eea

and

\bea
Z_{\a\b ,\c}& \equiv& \frac{1}{2}Y_{a(\a}{}^{\perp}T_{\b)\c}{}^a
-\frac{1}{2}Y_{\a\b}{}^{\d}h_{\d\c}+\frac{1}{2}Y_{\a\b,\c} \ ,
\nn\w2
Z_{a,\a\b}&\equiv&-\frac{1}{2}Y_{b\a}{}^{\perp}
(\C^{b})^{\c\d}\L_{a\c}h_{\d\b}+Y_{a\a}{}^{b}\L_{b\b} +
Y_{a\a}{}^{\d}h_{\d\b} -Y_{a\a,\b}\ ,
 \nn\w2
Z_{ab,\a}&\equiv&\frac{1}{2}Y_{ab,\a} -\frac{1}{2}Y_{ab}{}^{c}\L_{c\a}
-\frac{1}{2}Y_{ab}{}^{\c}h_{\c\a}\ .
\eea

It is also convenient to absorb all the perpendicular Dirac matrices by
defining the following objects (whose spinor indices can then be
treated as usual):

\bea
h_{\a\b}&\equiv& h_{\a}{}^{\b'}(\C^{\perp})_{\b'\b} \ ,\nn\w2
\L_{a\a}&\equiv& \L_{a}{}^{\a'}(\C^{\perp})_{\a'\a} \ ,\nn\w2 X_{\a
,\b\c}&\equiv& X_{\a ,\b}{}^{\c'}(\C^{\perp})_{\c'\c}\se
\frac{1}{2}(\C^{a})_{\b\c}X_{\a ,a}{}^{\perp} \ ,\nn\w2 X_{\a
,}{}^{\b\c}&\equiv& (\C^{\perp})^{\b\c'}X_{\a ,\c'}{}^{\c}\se
-\frac{1}{2}(\C^{a})^{\b\c}X_{\a ,a}{}^{\perp} \ ,\nn\w2 Y_{AB ,\c}
&\equiv& Y_{AB}{}^{\d'}(\C^{\perp})_{\d' \c}\ .
\eea

The tangential equations \eqs{T1}{T4} determine the worldvolume torsion
components in terms of the pull-backs of the target space torsion
components, the composite local connection $X_{a,b}{}^{\perp}$ and the
superfields $\L_{a\a}$ and $h_{\a\b}$.

Let us now examine the normal equations. At dimension zero we have the
algebraic equation \eq{A} which plays an important role in determining
the worldvolume supermultiplets and in solving the remaining normal
equations. At dimension half, eq. \eq{B} relates the 'hooked' component
of the spinorial derivative to the same structure in $Z_{\a\b,\c}$,
while the totally antisymmetric part $\nab_{[\a}h_{\b\c]}$ remains
undetermined (except in the cases of $p=1,2$, when this component
vanishes identically). Eq. \eq{B} also imposes the integrability
condition $Z_{(\a\b,\c)}=0$ on a curved target space background. Eq.
\eq{D} relates the local composite connection $X_{\a,a}{}^{\perp}$ to
the superembedding matrix element $\L_{a\a}$. At dimension one and
three half one can then demonstrate that eq.\eq{C} and eq. \eq{F} are
identically satisfied. Finally the dimension one equation \eq{E} is
identically satisfied. In fact, this equation and eq. \eq{F} are
non-linear versions of the linear integrability conditions
$\partial_{[a}\partial_{b]}X^{c'}=0$ and
$\partial_{[a}\partial_{b]}\Th^{\b'}=0$. In summary, the torsion
equation leads to an unconstrained transverse Goldstone superfield with
supercovariant component expansion given by $\Th_\b|$, $h_{\a\b}|$,
$\nab_{[\a}h_{\b\c]}|$ up to order $\th^3$.

\pagebreak


\end{document}